\documentclass[journal]{vgtc}                     % final (journal style)
\onlineid{0}

\vgtccategory{Research}

\vgtcpapertype{Technique}

\title{Probabilistic Inclusion Depth for Fuzzy Contour\\ Ensemble Visualization}

\author{%
\authororcid{Cenyang Wu}{0009-0001-0675-7317}, \authororcid{Daniel Kl{\"o}tzl}{0000-0002-4222-3320}, \authororcid{Qinhan Yu}{0009-0004-0445-0786}, \authororcid{Shudan Guo}{0009-0001-8395-6643},  \authororcid{Runhao Lin}{0009-0009-5550-0321}, \\ \authororcid{Daniel Weiskopf}{0000-0003-1174-1026}, and \authororcid{Liang Zhou*}{0000-0002-0462-4131}
}

\authorfooter{
  \item
  	Cenyang Wu and Liang Zhou are with the Institute of Medical Technology, Peking University Health Science Center and National Institute of Health Data Science, Peking University.
  	E-mails: wuceny@stu.pku.edu.cn, zhoulng@pku.edu.cn.
    Liang Zhou is the corresponding author.
    \item
  	Daniel Kl{\"o}tzl and Daniel Weiskopf are with the Visualization Research Center (VISUS), University of Stuttgart.
  	E-mails: firstname.lastname@visus.uni-stuttgart.de.
    \item
  	Qinhan Yu is with the Center for Machine Learning Research, Peking University.
  	E-mail: yuqinhan@stu.pku.edu.cn.
    \item
  	Shudan Guo and Runhao Lin are with the National Institute of Health Data Science, Peking University and Shandong University (SG), and Beijing Forestry University (RL).
  	E-mails: 202300300301@mail.sdu.edu.cn (SG), and meruem@bjfu.edu.cn (RL)
    
}

\abstract{%
We propose Probabilistic Inclusion Depth (PID) for the ensemble visualization of scalar fields.  
By introducing a probabilistic inclusion operator $\subset_{\!p}$, our method is a general data depth model supporting ensembles of fuzzy contours, such as soft masks from modern segmentation methods, and conventional ensembles of binary contours.
We also advocate for extending contour extraction in scalar field ensembles to become a fuzzy decision by considering the probabilistic distribution of an isovalue to encode the sensitivity information. 
To reduce the complexity of the data depth computation, an efficient approximation using the mean probabilistic contour is devised.
Furthermore, an order-of-magnitude reduction in computational time is achieved with an efficient parallel algorithm on the GPU.
Our new method enables the computation of contour boxplots for ensembles of probabilistic masks, ensembles defined on various types of grids, and large 3D ensembles not studied by existing methods.
The effectiveness of our method is evaluated through numerical comparisons with existing techniques on synthetic datasets, examples of real-world ensemble datasets, and expert feedback.
}


\teaser{\vspace{1.5em}
    \includegraphics[width=\textwidth]{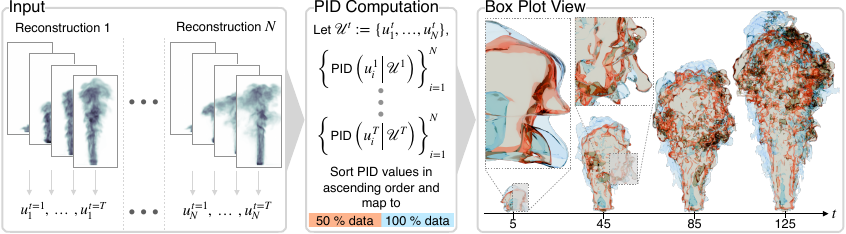}
    \caption{%
    The process of our PID method is applied to a 3D smoke plume ensemble of $N$ reconstructions. The PID is computed for each time step of the reconstructions with a GPU-accelerated PID-mean variant to create visualizations of contour boxplots that summarize the many members of delicate shapes.
    Dark regions of 3D boxplots are due to complex refractions.
    }
  \label{fig:teaser}
}

\graphicspath{{figs/}{figures/}{pictures/}{images/}{./}} % where to search for the images

\usepackage{tabu}                      % only used for the table example
\usepackage{booktabs}                  % only used for the table example
\usepackage{lipsum}                    % used to generate placeholder text
\usepackage{mwe}                       % used to generate placeholder figures
\usepackage{amsfonts}
\usepackage{amsmath}
\usepackage{graphicx}
\usepackage{svg} 
\usepackage{float}
\usepackage{comment}
\usepackage{empheq}
\usepackage{mathptmx}                  % use matching math font
\usepackage{color}
\usepackage{algorithm}
\usepackage{algpseudocode}

\newcommand{\INinp}{\ensuremath{\mathrm{IN}_{\mathrm{in}}^{\mathrm{p}}}}
\newcommand{\INoutp}{\ensuremath{\mathrm{IN}_{\mathrm{out}}^{\mathrm{p}}}}
\newcommand{\INin}{\ensuremath{\mathrm{IN}_{\mathrm{in}}}}
\newcommand{\INout}{\ensuremath{\mathrm{IN}_{\mathrm{out}}}}

\begin{document}

%%%%%%%%%%%%%%%%%%%%%%%%%%%%%%%%%%%%%%%%%%%%%%%%%%%%%%%%%%%%%%%%
%%%%%%%%%%%%%%%%%%%%%% START OF THE PAPER %%%%%%%%%%%%%%%%%%%%%%
%%%%%%%%%%%%%%%%%%%%%%%%%%%%%%%%%%%%%%%%%%%%%%%%%%%%%%%%%%%%%%%%

%% First section / Title -------------------------------------------------------
\firstsection{Introduction}
\maketitle

This paper addresses the issue of ensemble visualization of scalar field data and their sets of contours by relying on data depth.
In general, statistical depth functions provide a center-outward ordering of elements in a dataset---with the advantage of being non-parametric and robust against distortions from outliers. Therefore, they serve as the basis for non-parametric visualization methods like contour boxplots~\cite{whitakerContourBoxplotsMethod2013}. 

However, existing data depth models are limited to binary masks, precluding their use on contours with uncertainty or fuzzy contours. 
Examples of fuzzy contours include the increasingly common \emph{probabilistic masks} (or soft masks) produced by modern deep learning models (e.g., softmax outputs in medical image segmentation) or by probabilistic forecasting methods in meteorology.
Furthermore, existing data depth formulations for contours and scalar fields tend to be computationally intensive.
Early approaches such as Contour Band Depth (CBD) involve high algorithmic complexity~\cite{whitakerContourBoxplotsMethod2013}, whereas the more recent Inclusion Depth (ID) provides a more efficient and geometrically intuitive alternative by measuring inclusion relationships~\cite{chaves-de-plazaInclusionDepthContour2024}.
However, these methods are still not efficient enough for large 2D or 3D contour ensembles.

This work aims to bridge this gap. Our goal is to develop depth computation methods that handle uncertainty while simultaneously reducing computational complexity, making them practical for large-scale, high-resolution 3D ensemble data.
Note that in this paper, we use ``contour'' as a general term for dimensionally agnostic spatial domains, and use ``contour'' and ``surface'' interchangeably for 3D cases. 

Our method, Probabilistic Inclusion Depth (PID), is based on a probabilistic inclusion operator that describes the inclusion between two probabilistic masks.
This operator yields many desirable properties and carries over to PID, for example, Epsilon Inclusion Depth (eID)~\cite{chaves-de-plazaInclusionDepthContour2024} is a binary specialization of PID, and PID is coordinate-agnostic.
To make the processing of 3D data feasible, we devise an efficient approximation to PID and provide a parallel algorithm for GPU acceleration.
Our method is evaluated with ranking consistency tests and scalability studies comparing to existing data depth methods.
The usefulness of our method is demonstrated using real-world 3D ensembles from various applications, and through feedback from experts.

An example of our method applied to a 3D smoke plume ensemble~\cite{ScalarFlow2019} is shown in \cref{fig:teaser}.
For each time step, our method takes the 3D probabilistic maps modeled from each scalar field member (\cref{fig:teaser}--Input).
Depth values by PID are calculated for all input members, and then sorted and mapped to different percentiles of the population (\cref{fig:teaser}--PID Computation).
Subsequently, contour boxplots are generated by the unions and intersections of surfaces of the grouped members (\cref{fig:teaser}--Boxplot View).
Here, contour boxplots of four time steps are visualized (orange for 50\% population and blue for 100\% population) to show the progression of the smoke.

These are our main contributions:
\begin{itemize}
    \item We introduce a probabilistic inclusion operator ($\subset_{\!p}$) and Probabilistic Inclusion Depth (PID), enabling depth computation on fuzzy contours.
    \item We propose PID-mean, which reduces the computational complexity of PID to linear.
    \item A parallel algorithm further accelerates depth computation on the GPU, making the processing of complex 3D datasets feasible.
\end{itemize}
The source code is publicly available at \url{https://github.com/cenyangWu/Probabilistic-Inclusion-Depth}.
%% ---------------------------------------------------------------------------
\section{Related Work}

We discuss related work in the general context of ensemble visualization and data depth, followed by more specific techniques for contour depth and uncertainty of contours and surfaces.

\subsection{Ensemble Visualization and Data Depth}

Ensemble visualization is an active research topic~\cite{obermaierFutureChallengesEnsemble2014} and a subset of uncertainty visualization.
A comprehensive survey of ensemble visualization can be found elsewhere~\cite{Wang2019}.
A group of methods uses order statistics based on the data depth concept.
Data depth is a statistical concept~\cite{dyckerhoffDataDepthsSatisfying2004} originally used to describe the order statistics of multivariate data points.
Several multivariate data depth notions are available~\cite{mahalanobis1936generalized,Tukey1974,Oja1983,Liu1990}.
Band depth~\cite{Lopez-PintadoBandDepth2009} was introduced to characterize functions with a center-outward ordering.
An extension of band depth allows the creation of functional boxplots~\cite{sunFunctionalBoxplots2011}.
Surface boxplots are an extension of functional boxplots for summarizing ensembles of 2D images~\cite{Genton2014}. 
Half-region depth using the hypograph and epigraph of a curve is available for efficient depth computation for functions~\cite{lopez-pintadoHalfregionDepthFunctional2011}.

\subsection{Contour Depth Methods}

In visualization, the notion of data depth is extended to support ensembles of contours and curves.
With an extended band depth definition, order statistics of contours~\cite{whitakerContourBoxplotsMethod2013} and curves~\cite{mirzargarCurveBoxplotGeneralization2014a} can be calculated to generate boxplot visualizations for these features in 2D and even 3D~\cite{rajEvaluatingShapeAlignment2016}.
Localized band depth computation allows for the computation of representative consensus for limited-size ensembles~\cite{mirzargarRepresentativeConsensusLimitedsize2018}.
The main building block is a model of contour depth statistics to quantify the centrality of a shape within an ensemble. 

Unfortunately, the classic method for computing contour depth statistics---the Contour Band Depth (CBD)~\cite{whitakerContourBoxplotsMethod2013}---is relatively slow, making it impractical for large datasets. 
The recent approaches of Inclusion Depth (ID) and Epsilon Inclusion Depth (eID)~\cite{chaves-de-plazaInclusionDepthContour2024} reduce the computational complexity of depth computation. 
The concept of Inclusion Depth is based on the extension of half-region depth to contours, and can further be extended to support multi-modal contours~\cite{chaves-de-plazaDepthMultimodalContour2024}.
These methods, however, are designed for binary contours and cannot directly process probabilistic information in contours. 

\subsection{Uncertainty in Contours and Surfaces}

The uncertainty or fuzziness of contours and surfaces is studied in visualization.
The positional uncertainty of isocontours in a random field can be quantified and visualized using numerical condition~\cite{pothkowPositionalUncertaintyIsocontours2011}.
Probabilistic marching cubes generalize the uncertain isocontour model to spatially correlated random fields~\cite{pothkowProbabilisticMarchingCubes2011}.
An extension of probabilistic marching cubes takes distance-dependent correlations into account~\cite{Pfaffelmoser2011}.
Another extension uses a non-parametric model to characterize the uncertain scalar field~\cite{Athawale2016}.
The behavior of interpolation can be used within probabilistic marching cubes~\cite{Athawale2013}.
Uncertainty visualization is also available for fibers extracted from uncertain bivariate fields~\cite{Athawale2023}.

In contrast to these works, where the isocontours are modeled in uncertain scalar fields, we study fuzzy contours extracted from scalar fields without uncertainty by using a range instead of a single isovalue for robust and sensitivity visualization of ensembles.
This range modeling is in line with interval volumes devised as a generalization of isosurfacing~\cite{Fujishiro1995,Fujishiro1996,Ament:2010:DirectIntervalVolume}.
An interval volume is extracted from a volume using a continuous range of scalar values and describes generalized isosurfaces with a finite extent in the spatial domain for visualizing regions of uncertainty.

Another important application area of our method comes from modern segmentation techniques.
These methods, particularly deep neural networks, often output soft masks where each pixel value represents a probability or confidence score~\cite{ronneberger2015u,chen2017deeplab,zheng2021rethinking,kirillov2023segment}.
Techniques like Monte Carlo Dropout (MC-Dropout)~\cite{kendall2017uncertainties}, Deep Ensembles~\cite{lakshminarayanan2017simple}, and Temperature Scaling~\cite{guo2017calibration} are used to estimate this uncertainty.

Methods are available for comparing contours and surfaces. 
Isosurface Similarity Maps (ISM)~\cite{Bruckner2010,8440120} convert binary isosurfaces into continuous distance fields for similarity computation, but cannot directly process fuzzy contours.
While metrics like fuzzy Dice~\cite{ye2018generalized} or probabilistic IoU~\cite{lin2017refinenet} can be used to compare soft masks, and ISM for binary contours, they do not provide the geometric, center-outward ordering of statistical depth.

\section{Probabilistic Inclusion Depth}
We formulate Probabilistic Inclusion Depth, discuss its properties, and delineate it from the (Epsilon) Inclusion Depth.

\subsection{Background of Epsilon Inclusion Depth}
Our data depth model extends and generalizes Inclusion Depth (ID) and Epsilon Inclusion Depth (eID). Here, we briefly summarize the previous work by Chaves-de-Plaza et al.~\cite{chaves-de-plazaInclusionDepthContour2024}. We follow their description that aims at the 2D case, but with the understanding that the 3D case would be analogous.

The concept of ID is based on the containment information of contours for data depth computation. The extension to eID uses a continuous subset operator  $\subset_{\epsilon}$ for two sets $A, B \subset \mathbb{R}^2$ to compute the containment information: 
\begin{equation}
A \subset_{\epsilon} B = 1 - 
\begin{cases} 
0 & \text{if } |A| = 0\;, \\
\frac{|A \setminus B|}{|A|} & \text{otherwise}\;,
\end{cases}
\end{equation}
where $|A|$ denotes the area of set $A$ and $A \setminus B$ is the set difference. This operator returns a value in $[0, 1]$, which is 1 if $A \subset B$ and decreases as more of $A$ lies outside of $B$.

Let us now consider an ensemble of contours $\mathcal{C} = \{c_1, c_2, \dots c_N\}$, where each contour $c_i$ is associated with a scalar field $F_i\colon \Omega \to \mathbb{R}$ and an isovalue $q_i$.
The scalar fields $F_i$ map spatial locations $x \in \Omega\subset  \mathbb{R}^2$ to scalar values. 
The region inside a contour $c_i$ is then defined as:
\begin{equation}
    \text{in}(c_i) = \{x \; | \; F_i(x) < q_i\}\;.
\end{equation}
The definition of eID uses the continuous subset operator along with the notion of an inside area by first computing fractions of inclusions: 
\begin{equation}
\begin{aligned}
\INin(c_i) &= \frac{1}{N} \sum_{j=1}^{N} \mathrm{in}(c_i) \subset_{\epsilon} \mathrm{in}(c_j)\;, \\
\INout(c_i) &= \frac{1}{N} \sum_{j=1}^{N} \mathrm{in}(c_j) \subset_{\epsilon} \mathrm{in}(c_i)\;.
\end{aligned}
\end{equation}
Then, eID is the minimum of these two values:
\begin{equation}
\mathrm{eID}(c_i|\mathcal{C}) = \min\{\INin(c_i), \INout(c_i)\}\;.
\end{equation}
The same basic concept of containment serves as a basis for ID, which just replaces the continuous subset operator $\subset_{\epsilon}$ with the traditional $\subset$.

\subsection{Probabilistic Inclusion Depth Formulation}

All prior contour depth models rely on a binary decision: whether a part of the domain (e.g., a pixel or voxel) lies inside or outside the contour. We want to generalize this concept to a gradual membership (i.e., a gradual measure of being inside or outside the contour). To this end, we introduce a probabilistic or fuzzy mask $u:\Omega\!\to\![0,1]$, where $u$ maps each spatial location to a probability value. Therefore, instead of a curve that defines a sharp contour line, we now have an implicit representation by a field of probability values that indicates for each spatial location the gradual membership to the interior of the fuzzy~contour.
\begin{figure}[tb]
    \centering
    \includegraphics[width=\linewidth]{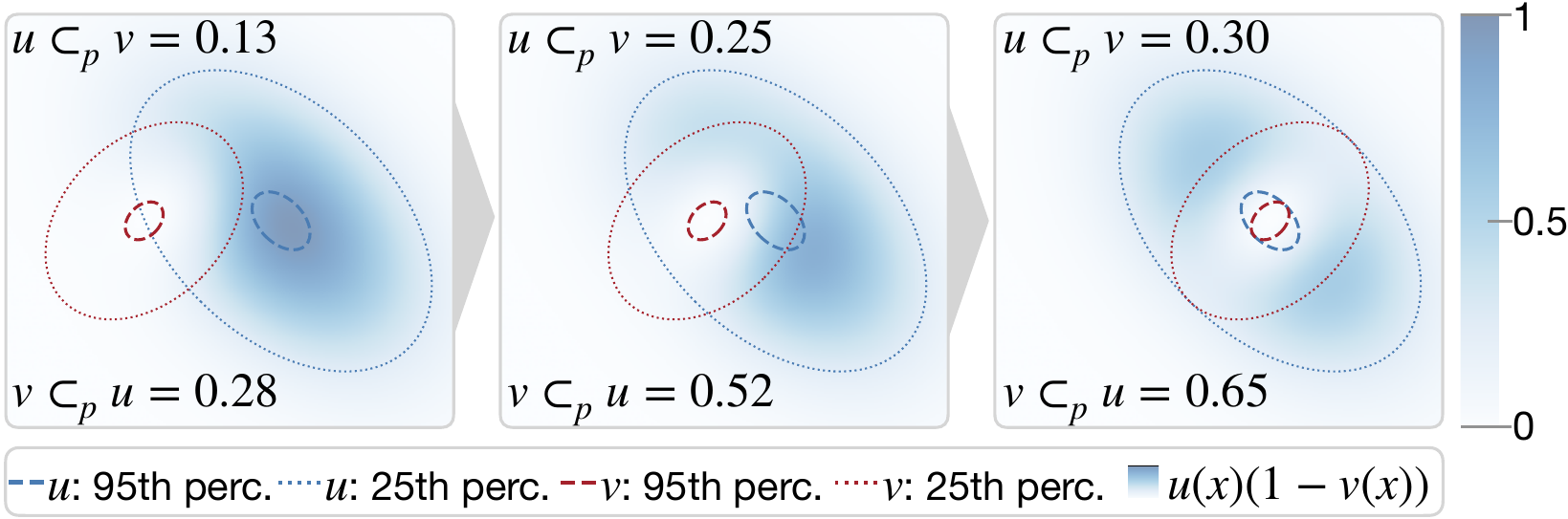}
    \caption{Visualization of the probabilistic inclusion operator applied to two 2D Gaussian distributions. 
    The fixed blue Gaussian $u$ is shown with covariance ellipses at $25\%$ and $95\%$ quantiles, while the red Gaussian $v$ moves along the horizontal axis. 
    The heatmap represents the element-wise penalty term $u(x)(1-v(x))$ of $u\subset_{\!p}v$.}
    \label{fig:probabilistic_inclusion_operator}
\end{figure}

We define inclusion for two probabilistic masks $u$ and $v$ as the expectation of $v$ with respect to the probability measure $\pi_u$ induced by $u$.
In this formulation, we work with probabilistic masks whose pixel/voxel values represent probabilities in $[0,1]$, and introduce a probabilistic inclusion operator $\subset_{\!p}$ for handling uncertainty in contour analysis. \Cref{fig:probabilistic_inclusion_operator} illustrates the probabilistic inclusion operator for several configurations of Gaussians $u$ and $v$.

The following description targets a general setup, working for any finite dimensionality of the domain for both continuous and discrete, grid-based domains. 
The key insight is to measure inclusion probabilistically.
Our probabilistic inclusion operator generalizes eID's $\subset_\epsilon$ by extending the set difference from $|A\setminus B|$ for binary sets to $u(1-v)$ for probabilistic masks, and extending the computation from counting grid points to accumulating the probabilities. 
When $u=\mathbf{1}_A$ and $v=\mathbf{1}_B$ are binary masks (indicator functions for sets $A$ and $B$), the probabilistic inclusion operator $\subset_{\!p}$ reduces to:
\begin{equation}
u\subset_{\!p}v
=\frac{\mu(A\cap B)}{\mu(A)}
=1-\frac{\mu(A\setminus B)}{\mu(A)}
= A\subset_{\epsilon} B\;,
\label{eq:binary_reduction}
\end{equation}
which coincides exactly with eID's $\subset_\epsilon$.

\paragraph{Definition via Expectation Under Induced Measure}
Let the measure space $(\Omega,\mathcal{F},\mu)$ be the mathematical basis for our modeling, where $\Omega$ represents the spatial domain (e.g., a 2D image or 3D volume), $\mathcal{F}$ a $\sigma$-field of measurable sets, and $\mu$ a measure (either counting measure for discrete grids or Lebesgue measure for continuous domains). The spatial location (pixel or voxel coordinate) is denoted as $x \in \Omega$.

For a probabilistic mask $u:\Omega\!\to\![0,1]$ (where $u$ maps each spatial location to a probability value), we define its mass as
\begin{equation}
m(u)=\int_\Omega u(x)\,\mathrm d\mu(x)\;,
\end{equation}
i.e., the total probabilistic ``volume'' of the mask.

With this, we arrive at a probability measure $\pi_u$ induced by $u$, which reads for any measurable set $E \subseteq \Omega$: 
\begin{equation}
\pi_u(E)=\frac{1}{m(u)}\int_E u(x)\,\mathrm d\mu(x)
\;,\quad 
\frac{\mathrm d\pi_u}{\mathrm d\mu}(x)=\frac{u(x)}{m(u)}\quad(m(u)>0)\;.
\end{equation}
Now, we can define the probabilistic inclusion operator as
\begin{equation}
u\subset_{\!p}v \;:=\; \mathbb{E}_{X\sim \pi_u}\![\,v(X)\,]
\;=\; \int \frac{u(x)}{m(u)}\,v(x)\,\mathrm d\mu(x)\;,
\label{eq:expectation_form}
\end{equation}
where $\mathbb{E}[\cdot]$ denotes the expectation and $X$ is a random variable distributed according to $\pi_u$ (i.e., $X\sim \pi_u$). Equivalently, by change of measure, we also have $\mathrm d\pi_u/\mathrm d\mu=u/m(u)$. We adopt the convention $u\subset_{\!p}v=0$ if $m(u)=0$, i.e., when the reference mask has zero mass.

By a single algebraic step, we can reformulate the operator as 
\begin{equation}
u \subset_{\!p} v = \mathbb{E}_{X\sim \pi_u}[v(X)]
=\frac{\int u\,v\,\mathrm d\mu}{\int u\,\mathrm d\mu}
=1-\frac{\int u(1-v)\,\mathrm d\mu}{\int u\,\mathrm d\mu}\;.
\label{eq:prob_contain}
\end{equation}

\paragraph{Interpretation and Semantics}
The expectation-under-$\pi_u$ view provides an interpretable inclusion semantics: sample locations in the spatial domain according to $u$ (i.e., more likely locations where the probability $u$ is high) and take the weighted average of $v$ then. 
It naturally captures directionality ($u\subset_{\!p}v \neq v\subset_{\!p}u$), attains intuitive limiting cases (if $v=1$ $\mu$-a.e. on $\{u>0\}$, the value is $1$; if $v=0$ $\mu$-a.e., it is $0$), reduces to set inclusion for binary masks (as shown in \cref{eq:binary_reduction}), and for probabilistic masks yields values below $1$ that quantify how much this uncertainty diminishes inclusion.

\paragraph{Definition of Probabilistic Inclusion Depth}
Using the probabilistic inclusion operator $\subset_{\!p}$, we define PID for an ensemble of $N$ contours $\{c_1, c_2, \ldots, c_N\}$, where each contour $c_i$ is represented by its corresponding probabilistic mask $u_i$:
\newcommand*\widefbox[1]{\fbox{\hspace{2em}#1\hspace{2em}}}
\begin{equation}
\begin{aligned}
& \INinp(c_i) = \frac1N\sum_{j=1}^{N}\!\bigl(u_i\subset_{\!p}u_j\bigr)\;, \quad
  \INoutp(c_i) = \frac1N\sum_{j=1}^{N}\!\bigl(u_j\subset_{\!p}u_i\bigr)\;, \\
&\qquad\qquad\mathrm{PID}(c_i) = \min\!\bigl\{\INinp(u_i),\,\INoutp(u_i)\bigr\}\;. 
\end{aligned}
\label{eq:pid}
\end{equation}
Here, $\INinp(c_i)$ represents how much contour $c_i$ is contained within other contours in the ensemble (average inclusion of $c_i$ by others), whereas $\INoutp(c_i)$ describes how much $c_i$ contains other contours (average inclusion of others by $c_i$). 
The final PID value of $c_i$ is defined as the minimum of these two quantities.

\begin{figure}[tb]
    \centering
    \includegraphics[width=0.95\linewidth]{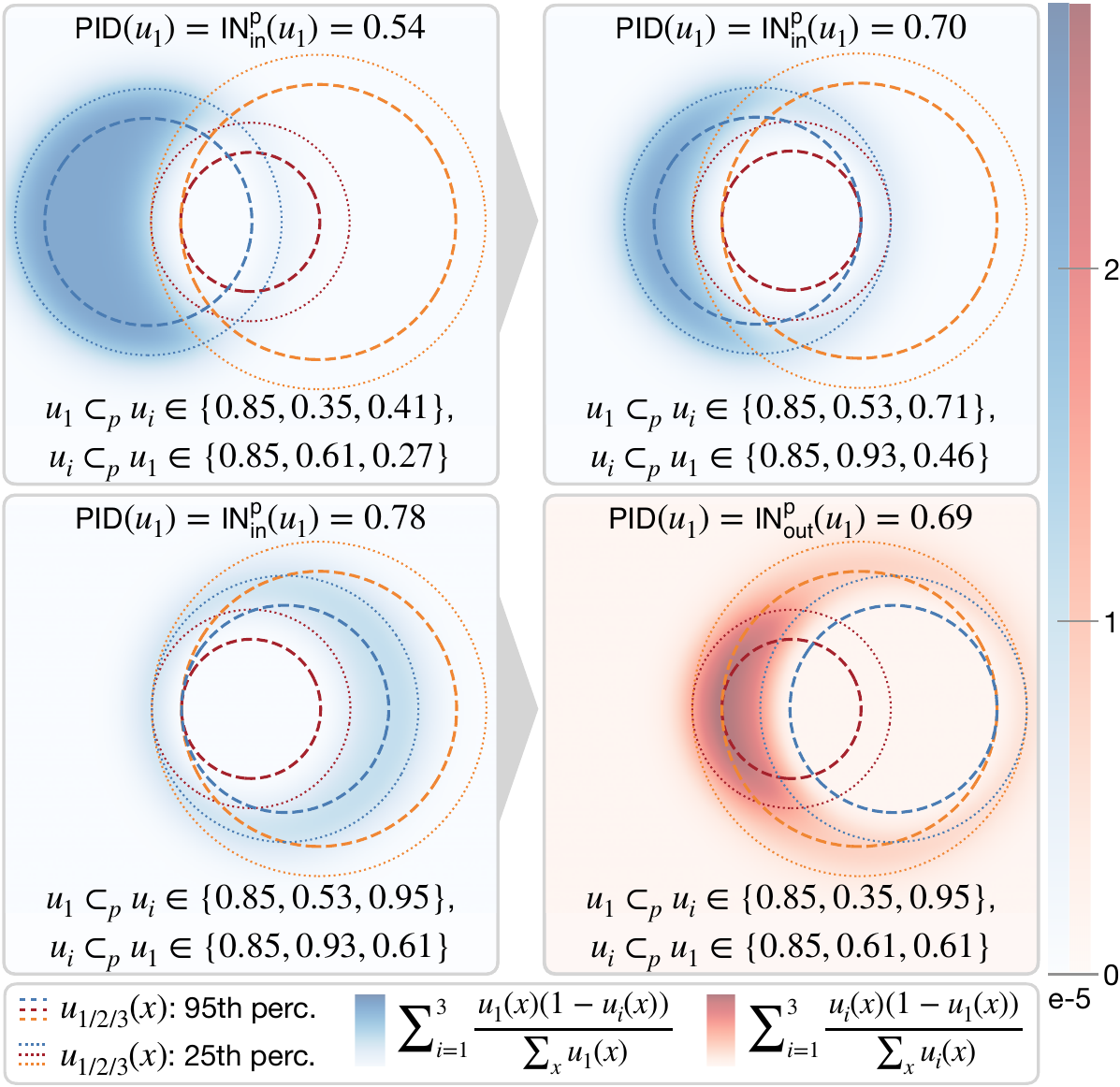}
    \caption{Probabilistic inclusion between three fuzzy (soft) disks. Here, the blue disk $u_1$ moves horizontally while the red ($u_2$) and orange ($u_3$) disks are fixed. Density values of each disk are 1 inside given radii (3, 2, and 4) and decay normally distributed, with fixed variance 0.8. The computation formulas for $\INinp$ and $\INoutp$ are given in \cref{eq:pid}. Depending on which of the two is minimal for $u_1$, the overlaid heatmap represents the respective element-wise penalty term for the probabilistic inclusion depth.}
    \label{fig:PID_rings}
\end{figure}

\Cref{fig:PID_rings} illustrates this computation using three fuzzy (soft) disks. 
It shows how the PID value $\textrm{PID}(u_1)$ and corresponding inclusion values ($u_1 \subset_{\!p} u_i$ and $u_i \subset_{\!p} u_1$) change when shifting the probabilistic mask $u_1$ from left to right. 
The overlaid heatmaps visualize the element-wise contribution to the penalty term from \cref{eq:prob_contain}, which underlies either $\INinp(u_1)$ or $\INoutp(u_1)$, depending on which of the two is minimal. 
High intensity indicates regions where mismatches between $u_1$ and the other masks strongly contribute to lowering the inclusion value, and hence to smaller PID. 
In this example, the distinction between $\INinp$ and $\INoutp$ becomes visible in the two lower plots. 
On the left, $u_1$ is centered between $u_2$ and $u_3$, which leads to high outer inclusions $u_i\subset_{\!p}u_1$ and thus $\INoutp(u_1)$ dominates. On the right, as $u_1$ moves away from $u_2$, the outer inclusions decrease and $\INoutp(u_1)$ drops below $\INinp(u_1)$, making $\INoutp$ the determining factor for $\textrm{PID}(u_1)$.

\subsection{Properties}
\label{sec:properties}

The probabilistic inclusion operator and PID have many useful properties. Details of these are discussed in the supplemental material~\cite{supplemental:ProbInclDepth}. In this section, we restrict ourselves to a short summary.

Let us start with a list of observations for the probabilistic inclusion operator $u \subset_{\!p} v$. It is linear in $v$, monotonic in $v$, scale-invariant in $u$, asymmetric (directional), and Lipschitz-continuous both in $u$ and $v$.

As discussed above, for the binary specialization, the probabilistic inclusion operator is reduced to the $\subset_\epsilon$ operator of eID. In addition, for probabilistic masks that converge to binary masks, the result of the probabilistic inclusion operator converges to the binary specialization at a Lipschitz rate. Consequently, within the continuous-measure formulation of PID, eID arises as its binary special case.

The properties of the probabilistic inclusion operator carry over to PID. For example, PID is monotonic, it is consistent with binary specialization (i.e., PID reduces to eID when all masks are binary), and it is Lipschitz-continuous with respect to both $u$ and $v$ (i.e., robust under bounded perturbations).
Finally, PID is coordinate-agnostic: it depends only on the background measure $\mu$ over the domain $(\Omega,\mathcal F,\mu)$.
For any measurable bijection $T:\Omega\to\Omega'$ whose inverse is also measurable (a bi-measurable bijection), we define the transformed measure $\mu'$ on $\Omega'$ such that $\mu'(E) = \mu(T^{-1}(E))$ for any measurable set $E \subseteq \Omega'$, and we have:
\begin{equation}
\frac{\int_\Omega u\,v\,d\mu}{\int_\Omega u\,d\mu}
=\frac{\int_{\Omega'} (u\circ T^{-1})(v\circ T^{-1})\,d\mu'}{\int_{\Omega'} (u\circ T^{-1})\,d\mu'}\;.
\end{equation}
Consequently, the two PID terms
\(
\INinp=\frac1N\sum_{j=1}^{N}\![u_i\subset_{\!p}u_j]
\)
and
\(
\INoutp=\frac1N\sum_{j=1}^{N}\![u_j\subset_{\!p}u_i]
\)
are invariant under reparameterizations of the domain. Typical examples include translations and rotations in Euclidean space (with Lebesgue measure); on a Riemannian manifold, one can use the native volume measure
\(
\mathrm d\mu=\sqrt{\det g(x)}\,\mathrm dx\;,
\)
where $g(x)$ is the Riemannian metric tensor in local coordinates. 
This invariance under reparameterization is also the basis for arriving at the computational schemes for different spatial domain representations, eventually leading to include volume-weighted computations to reflect the measures over the domain.

\subsection{Typical Spatial Domain Representations}
\label{sec:spatialdomains}

Our description so far was designed to be abstract and generic. We now want to exemplify a few typical, specific cases of the spatial domain $\Omega$. 

\paragraph{Euclidean Space} The first case is that of a Euclidean space, typically of two or three dimensions. Here, coordinates $x$ are from this Euclidean space, and the measure $\mu$ is the Lebesgue measure (i.e., the usual measure of area or volume).

\paragraph{Uniform Grids}
Data in Euclidean space is often discretized on a uniform grid, i.e., a grid of pixels (2D) or voxels (3D). Assuming nearest-neighbor interpolation, the integration boils down to piecewise constant contributions, leading to 
\begin{equation}
u\subset_{\!p}v=1-\frac{\sum_x u(x)(1-v(x))}{\sum_x u(x)}\;.
\end{equation}
Here, the domain can be represented by a set of grid indices: $\Omega=\{1,\ldots, M\}$, where $M$ denotes the total number of pixels or voxels, and the $x$ correspond to pixel/voxel coordinates.

\paragraph{Nonuniform Grids}
Curvilinear grids or unstructured grids may contain cells of different shapes and sizes. Again, we assume cell-wise constant modeling. Let $w_x>0$ denote the physical area/volume of the cell associated with $x$. Then, we replace all sums by volume-weighted sums:
\begin{equation}
u\subset_{\!p}v= 1- \frac{\sum_x w_x\,u(x)\,(1-v(x))}{\sum_x w_x\,u(x)}\;.
\end{equation}
With this, the two PID terms become
\begin{equation}
\begin{aligned}
\INinp
&=\frac1N\sum_{j=1}^{N}\!\left[1-\frac{\sum_x w_x\,u_i(x)\,\bigl(1-u_j(x)\bigr)}{\sum_x w_x\,u_i(x)}\right]\;, \\
\INoutp
&=\frac1N\sum_{j=1}^{N}\!\left[1-\frac{\sum_x w_x\,u_j(x)\,\bigl(1-u_i(x)\bigr)}{\sum_x w_x\,u_j(x)}\right]\;.
\end{aligned}
\end{equation}

The computation for a uniform grid is a special case in which all weights are equal.

\begin{figure}[b]
    \centering
    \includegraphics[width=\linewidth]{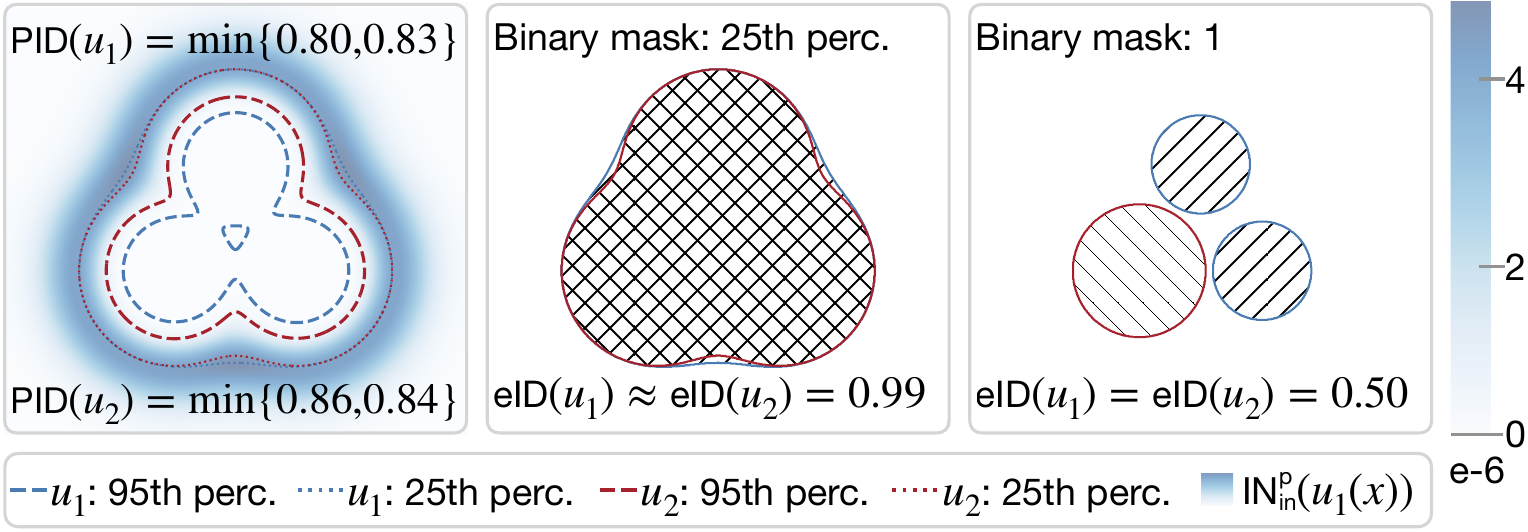}
    \caption{The left panel shows two highly similar Gaussian-blurred disk ensembles, $u_1$ (blue) and $u_2$ (red), with reference contours at the 25\% and 95\% quantiles. 
    The heatmap depicts the element-wise penalty term $\INinp(u_1(x))$. 
    The middle and right panels show binary masks at two threshold levels (the 25th percentile and the density value 1), resulting in different eID values. Hatched regions indicate support overlap.
    }
    \label{fig:PID_eID}
\end{figure}

\subsection{Modeling the Probabilistic Mask}
\label{sec:probabilisticmask}

The input probabilistic masks to PID can be modeled in several ways.
One approach is to use explicit soft masks such as those from segmentation techniques, as shown in~\cref{sec:brainTumor}.
The soft masks describe the probability of the appearance of object boundaries.
Soft masks from other machine learning techniques, such as attention maps of a transformer model, can also be used as input.

Another approach takes fuzzy contours of a scalar field ensemble by taking a probability distribution of the isovalue, as discussed in \cref{sec:compareEid} and \cref{sec:scalarFlow}.
This approach is similar to interval volumes~\cite{Fujishiro1995,Fujishiro1996,Ament:2010:DirectIntervalVolume} and can be used for visualizing the sensitivity of contouring in scalar~fields.

\begin{figure*}[tb]
    \centering
    \parbox[c]{.5\textwidth}
    {    
    \subfloat[Rank differences of eID]{\includegraphics[width=0.5\linewidth]{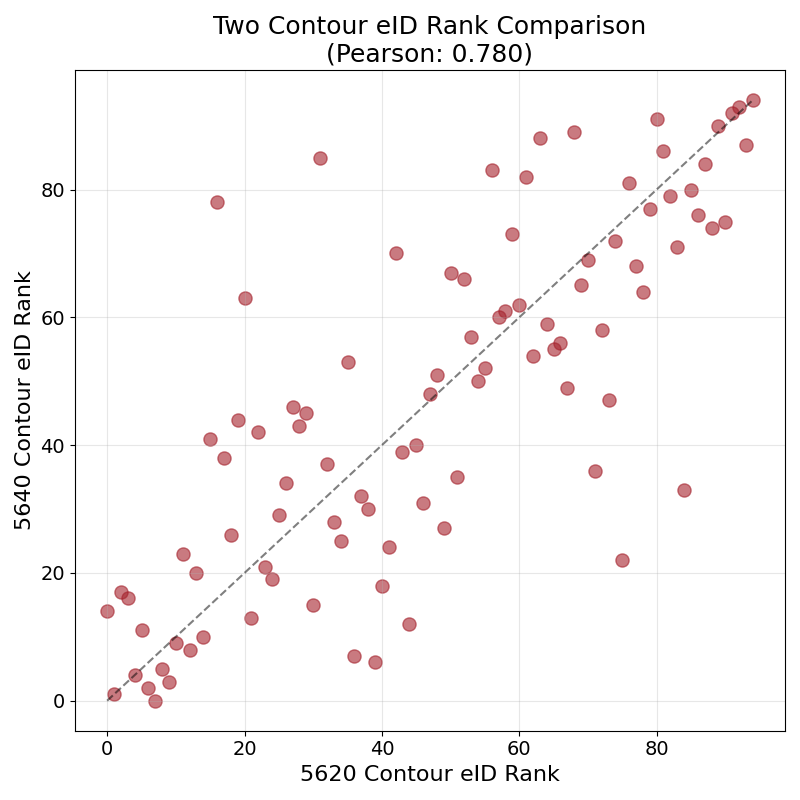}}
    \subfloat[Rank differences of PID]{\includegraphics[width=0.5\linewidth]{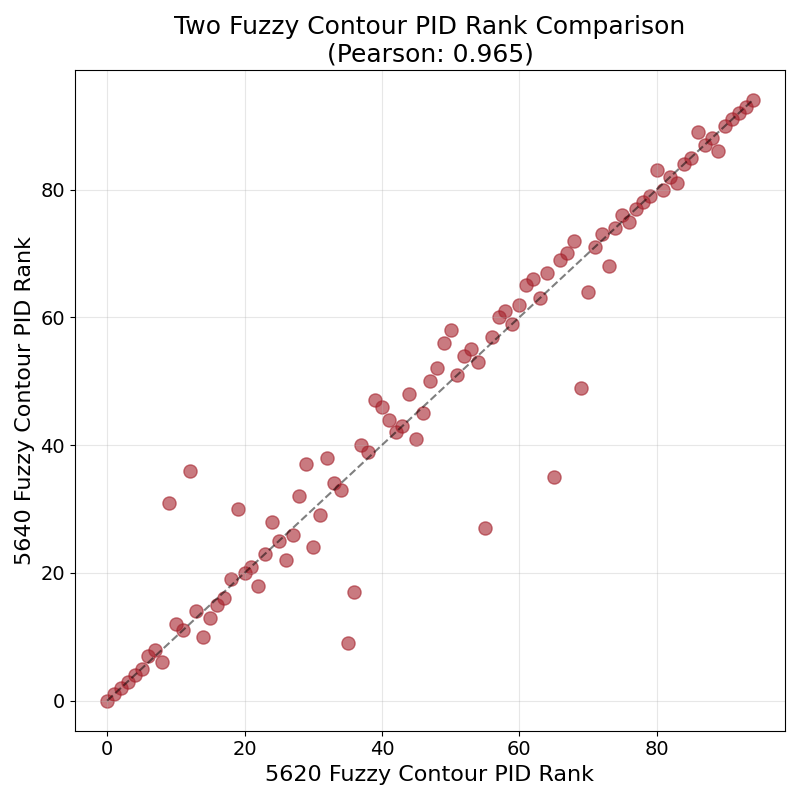}}
    }\hfill
    \parbox[c]{.49
    \textwidth}{
     \subfloat[Probabilistic maps of the member with highest PID rank differences]{\includegraphics[trim=0 2cm 0 1cm, clip, width=\linewidth]{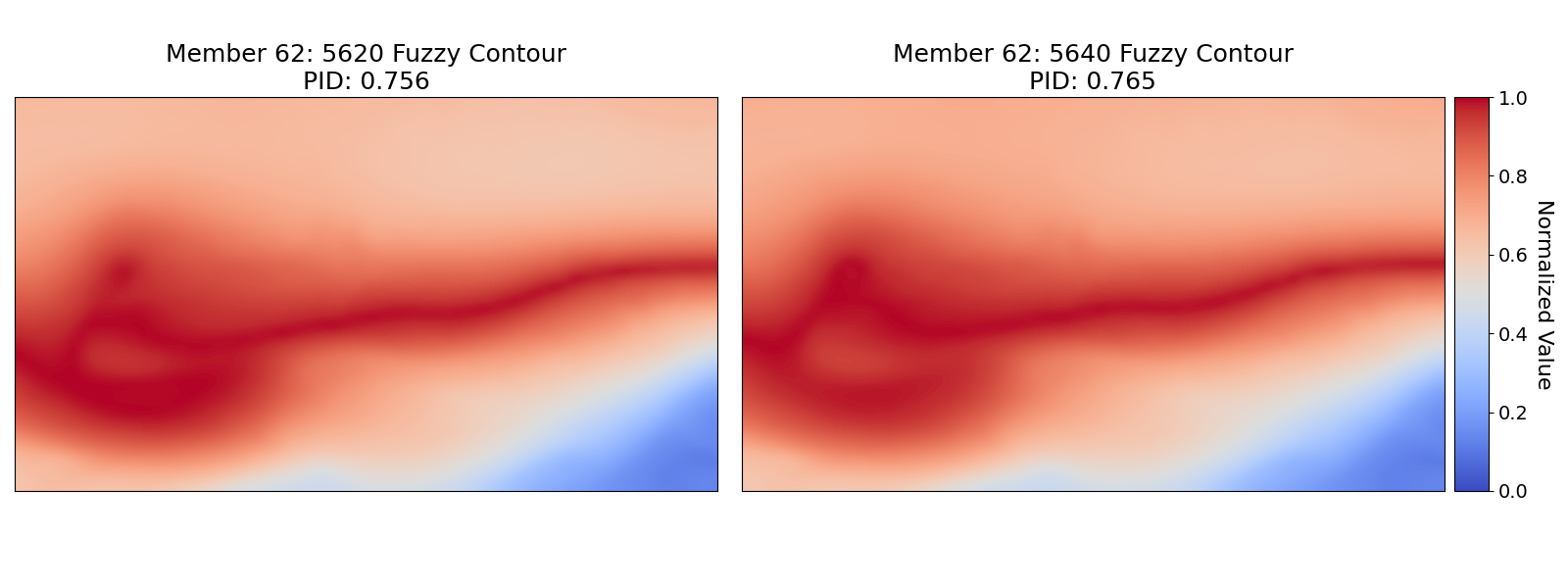}}\\
     \subfloat[Contours of the top 5 members with highest eID rank differences]{\includegraphics[trim=0 4cm 0 3cm, clip, width=\linewidth]{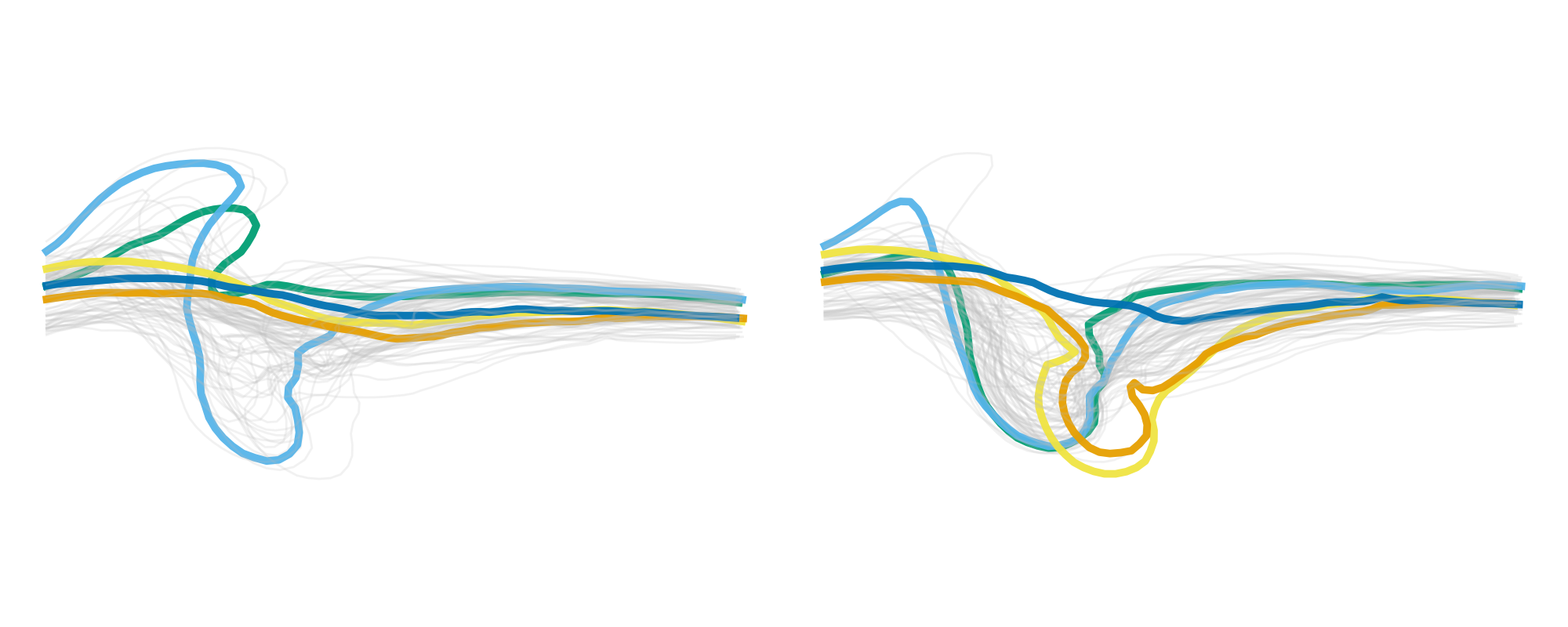}}
    }
    \caption{
    Comparisons of PID and eID on an ensemble of weather simulations. 
    The ranking differences are shown for (a) between eID-ranked binary contours of 5620\,m and 5640\,m, and (b) PID ranks of fuzzy contours centered at 5620\,m and 5640\,m. 
    Probabilistic maps of the member with the highest PID rank difference are visualized in (c), and (d) shows contours of the top 5 members with the highest eID rank differences. 
    }
    \label{fig:weather}
\end{figure*}
\subsection{Comparison to eID}
\label{sec:compareEid}
As discussed before, eID is a special case of PID for binary masks only. 
Here we highlight the differences between the two depth methods with~examples.
\paragraph{Illustrative Toy Example}
An illustrative example is shown in \cref{fig:PID_eID}. 
The example contains two uncertain regions ($u_1$, $u_2$) that are similar but not identical (\cref{fig:PID_eID}--left). 
A binary threshold at the 25\% percentile turns both regions into nearly identical masks, and yields a value of $0.99$ under eID, although the underlying probability fields differ (\cref{fig:PID_eID}--middle). 
A threshold of 1, by contrast, produces binary contours with markedly different shapes (\cref{fig:PID_eID}--right). 
In this case, the binary masks no longer coincide, and the eID depth values of both contours drop to $0.5$. 
This illustrates the sensitivity of eID to thresholding: the same pair of uncertain regions can either be collapsed into nearly identical ones or produce large geometric discrepancies with synchronized score drops, depending solely on the chosen threshold level. 
In contrast, PID is robust as it avoids thresholding and varies smoothly with the input probabilistic maps (\cref{fig:PID_eID}--left)---this behavior follows the Lipschitz continuity of PID (see \cref{sec:properties}).

\paragraph{Weather Simulation Ensemble}

In this weather simulation ensemble example (data created by the European Centre for Medium-Range Weather
Forecast\footnote{\url{https://www.ecmwf.int}}), we compare PID to eID at two height levels: 5620\,m and 5640\,m.
\Cref{fig:weather}(a) is the rank-to-rank scatterplot of 5620\,m versus 5640\,m when eID is applied to binary isocontours: the points show some deviation from the diagonal (Pearson $r=0.780$), indicating moderate rank changes under a 20\,m level shift. 
For PID, we construct fuzzy contours by mapping each target level (5620\,m and 5640\,m) to a level-centered probabilistic map where probabilities of~1 are assigned to the target value positions with a linear fall off to~0 using linear interpolation, i.e., a probabilistic distribution rather than a hard threshold. 
\Cref{fig:weather}(b) shows the comparison of PID ranks on the fuzzy contours centered at the two levels: points are concentrated near the diagonal (Pearson $r=0.965$), indicating much higher rank stability compared to eID. 
\Cref{fig:weather}(c) visualizes the two probabilistic maps for the member with the largest PID rank change between 5620\,m and 5640\,m---the heatmaps look very similar. 
Meanwhile, \cref{fig:weather}(d) plots the binary contours of the top five members with the largest eID rank changes, which differ. 
These observations indicate that binary thresholding decisions can induce sensitivity to small level shifts for eID, whereas probabilistic fields of fuzzy contours with PID yield smoother and more robust rank changes.

Importantly, PID provides insights that eID cannot: by operating directly on probabilistic data without binarization, PID captures the \emph{gradual transitions} and \emph{uncertainty magnitudes} inherent in the ensemble members, rather than forcing discrete inclusion/exclusion decisions at arbitrary thresholds. 
This enables PID to quantify how \emph{probabilistically similar} members are across different levels, revealing continuous variations in spatial patterns that eID's binary approach would conflate or miss. 
For instance, two members with slightly different probability distributions may yield identical binary contours under eID (thus identical ranks), but PID distinguishes their subtle differences, providing a finer-grained ranking that reflects the actual uncertainty structure.

\subsection{Comparison to Other Probabilistic Mask Similarities}

We compare PID with alternative depths using similarity measures that operate directly on probabilistic masks without thresholding, specifically, fuzzy Dice~\cite{ye2018generalized} and probabilistic IoU~\cite{lin2017refinenet}. 
Here, the depth of a member is calculated using the similarity measure between the member and the mean mask of the ensemble, and the rank is sorted by depth thereafter.
Probabilistic IoU is used for comparison, as we find that fuzzy Dice and probabilistic IoU exhibit a monotonic relationship and yield identical depth rankings.

\begin{figure}[htb]
   \centering
   \includegraphics[width=\linewidth]{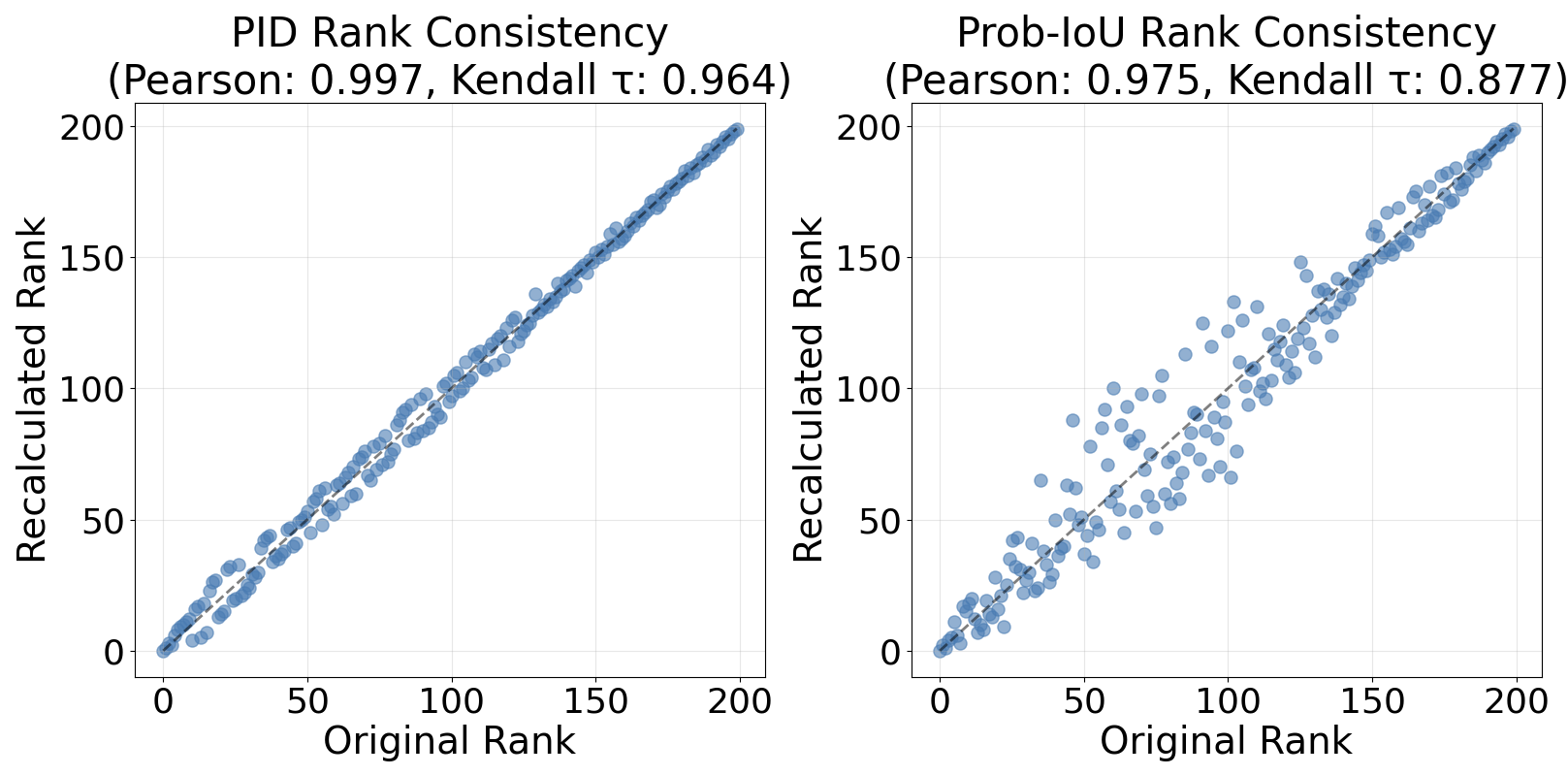}
   \caption{Ranking consistency comparison between PID and probabilistic IoU after removing the 10 lowest-ranked members.}
   \label{fig:IouConsistency}
\end{figure}

Using the synthetic 3D ellipsoid ensembles (\cref{sec:scalability}: 210 members--200 base + 10 extreme outliers, $50^3$ voxels), a stability test is performed: we rank all members first, and then remove the 10 lowest-ranked and recompute rankings for the remaining members. As shown in \cref{fig:IouConsistency}, PID has a higher ranking consistency according to Pearson correlation coefficients (0.997 vs.\ 0.975) and Kendall $\tau$ (0.964 vs.\ 0.877)---a correlation measurement by counting concordant and discordant pairs---and a better outlier detection capability compared to probabilistic IoU.

The difference lies in the pairwise inclusion analysis of PID, which captures geometric relationships between all ensemble members, whereas mean-based methods (probabilistic IoU/fuzzy Dice) compare each member only to the ensemble mean, making them susceptible to the influence of extreme values---particularly problematic for ensembles with high internal variability.

\paragraph{Summary of Advantages of PID}

In summary, PID offers the following advantages:

\begin{itemize}
\item \emph{Uncertainty-awareness without thresholding.} It operates directly on probabilistic maps (soft masks or normalized fields), preserving fuzzy boundaries and partial-volume/diffuse effects.
\item \emph{Robustness.} It is Lipschitz-continuous in both arguments: small input changes cause only small output changes, avoiding threshold-induced jumps.
\item \emph{Unified treatment.} It relies only on the background measure $\mu$ and can readily handle non-uniform grids; binary, probabilistic, and density fields are all valid inputs.
\end{itemize}

\section{Linear Complexity Probabilistic Depth Computation}
\label{sec:pidmean}
To overcome the quadratic complexity of PID, we introduce PID-mean, an approximation that reduces the complexity to be linear in the size of the ensemble.
Furthermore, a parallel algorithm is devised for GPU acceleration. In our discussion, we follow the characterization of visualization scalability~\cite{Richer:2024:Scalability} in terms of algorithmic complexity.

\subsection{PID-mean}

Instead of performing $N^2$ pairwise comparisons, PID-mean compares each contour $u_i$ only against a single mean contour $\bar{u}$ of the entire ensemble:
\begin{equation}
\bar u(x)=\frac1N\sum_{i=1}^{N}u_i(x)\;.
\end{equation}
It  then requires only two probabilistic inclusion calculations per contour:
\begin{equation}
\INin^{\text{mean}}(c_i)=u_i\subset_{\!p}\bar u,\quad \INout^{\text{mean}}(c_i)=\bar u\subset_{\!p}u_i\;.
\end{equation}
This reduces the overall complexity to $\mathcal{O}(N)$.
We show that PID-mean is a reasonable approximation of PID, as outlined in the rationale provided below.

\paragraph{Rationale of PID-mean}
Consider the ensemble of probabilistic contours as multiple delineations on a common spatial domain.
The superposition of these delineations produces a \emph{consensus intensity map} $\bar u$, where higher intensities indicate stronger agreement that a location belongs to the interior. 

For a given ensemble member $u_i$, PID-mean evaluates two complementary criteria and takes their minimum as the depth.
\begin{itemize}
    \item \textbf{The recall  aspect (member $\Rightarrow$ consensus)} measures the average consensus intensity $\bar u$ over locations typical for the member (weighted by $u_i$).
    \[
    u_i \subset_{\!p} \bar u = \mathbb{E}_{\pi_{u_i}}[\bar u]\;.
    \]
    \item \textbf{The precision aspect (consensus $\Rightarrow$ member)} describes the member probability $u_i$ on average over locations emphasized by the consensus (weighted by $\bar u$).
    \[
    \bar u \subset_{\!p} u_i = \mathbb{E}_{\pi_{\bar u}}[u_i]\;.
    \]
\end{itemize}
Taking the $\min$ operator acts as a conservative conjunction: only when both ``member$\Rightarrow$ consensus'' and ``consensus$\Rightarrow$ member'' are large is the contour appropriately \emph{central}. 
This construction avoids two types of spurious centers: \emph{large-but-shifted} members (broad support that overlaps the consensus yet is displaced overall; high recall but low precision) and \emph{small-but-peaky} members (covering only a small part of the consensus; high precision but low recall). 
Both are penalized by the $\min$ operator.

The mean consensus $\bar u$ is adopted for representativeness and efficiency. 
For the recall aspect, we have:
\begin{equation}
\label{eq:PIDmean}
 u_i \subset_{\!p} \bar u 
 \;=\; \mathbb{E}_{\pi_{u_i}}\!\Big[\tfrac1N\sum_{j=1}^N u_j\Big]
 \;=\; \tfrac1N\sum_{j=1}^N \mathbb{E}_{\pi_{u_i}}[u_j]
 \;=\; \tfrac1N\sum_{j=1}^N (u_i \subset_{\!p} u_j)\;,
\end{equation}
indicating that averaging the ensemble first and then comparing is equivalent to comparing pairwise and then averaging. 
Therefore, $\bar u$ provides a faithful group representative and allows for reducing the computational complexity. 
For the precision aspect, weighting by $\bar u$ furnishes a stable and interpretable representative assessment without the overhead of pairwise weighting; 
in practice, together with the recall aspect, it prevents situations of over-coverage or under-coverage, thereby yielding a robust and reasonable depth.

\paragraph{Error Bound} The approximation error $|\mathrm{PID}(u_i)-\mathrm{PID\text{-}mean}(u_i)|$ is bounded by the coefficient of variation (CV) of ensemble member masses. 
This bound ensures that PID-mean can be safely used when the ensemble exhibits low size variability (small CV), which is typical for ensembles generated from physical simulations or consistent segmentation networks.
For ensembles with high mass variability (large CV), the full pairwise PID should be preferred to avoid approximation artifacts.
Compared to fuzzy Dice or probabilistic IoU, although PID-mean also uses a mean-based approximation, it provides theoretical error bounds via the CV of ensemble masses, ensuring reliability when CV is low.
In practice, real-world ensembles typically have relatively low CV---supporting the use of PID-mean.
The complete mathematical derivation of the error bounds and the empirical CV values for all datasets used in this paper are provided in the supplemental material~\cite{supplemental:ProbInclDepth}.

\paragraph{Properties of PID-mean}

The mean probabilistic contour remains a point-wise (optionally weighted by reliability) average
that is independent of coordinates. 
Since the mean computation is coordinate-independent and only the integral operations depend on the domain measure, all the properties discussed in \cref{sec:properties} hold for both PID and PID-mean.

\subsection{Parallel Processing on the GPU}
To further accelerate PID-mean computation for large 3D ensembles, we devise a GPU-accelerated parallel algorithm using CUDA. 
The computation consists of per-voxel multiply–adds and global reductions, which are bandwidth-bound operations well-suited for GPU execution.
We map one ensemble member per thread block and perform binary-tree reductions in shared memory to accumulate the numerators and denominators of PID.
To maximize throughput, we employ coalesced memory access patterns.

Our method is implemented as a standalone preprocessing program that outputs the data depth, from either PID or PID-mean.
The detailed implementation and performance considerations are provided in the supplemental material~\cite{supplemental:ProbInclDepth}.

\section{Evaluations}
A ranking consistency test for its correctness and a scalability analysis for the computational performance were used to evaluate our method.
Experiments were conducted on a workstation with 32\,GB main memory, a 3.4\,GHz Intel i7 CPU, and an NVIDIA GeForce RTX 4080 Super GPU with 16 GB graphics memory, running on Windows~10.

\subsection{Ranking Consistency}
The ranking of ensemble members varies across different data depth methods, and a ground truth ranking is nonexistent.
Therefore, we evaluate the correctness of our PID-mean method by computing its ranking consistency against existing data depth methods using the Pearson correlation coefficient, which measures linear association between rankings.
A consistency analysis comparing PID-mean with eID (the binary specialization of PID), CBD, probabilistic IoU depth (Prob-IoU), and ISM depth (ISM) on a synthetic contour ensemble from Chavez-de-Plaza et al.~\cite{chaves-de-plazaInclusionDepthContour2024} was conducted.
CBD is a depth method that measures depth based on the frequency of being contained within bands formed by pairs of contours~\cite{whitakerContourBoxplotsMethod2013}.
Prob-IoU depth is computed as the probabilistic IoU~\cite{lin2017refinenet} similarity between each member and the ensemble mean.
Similarly, ISM depth is derived from the ISM similarity~\cite{Bruckner2010} between the member and the ensemble mean.
We computed depth rankings, where higher ranks indicate lower depth, for all ensemble members using these five methods on a dataset containing 200 generated contours. 

\begin{figure}[tb]
%\vspace{-1em}
    \centering
    \includegraphics[width=0.9\linewidth, trim=5cm 4cm 5cm 5cm, clip]{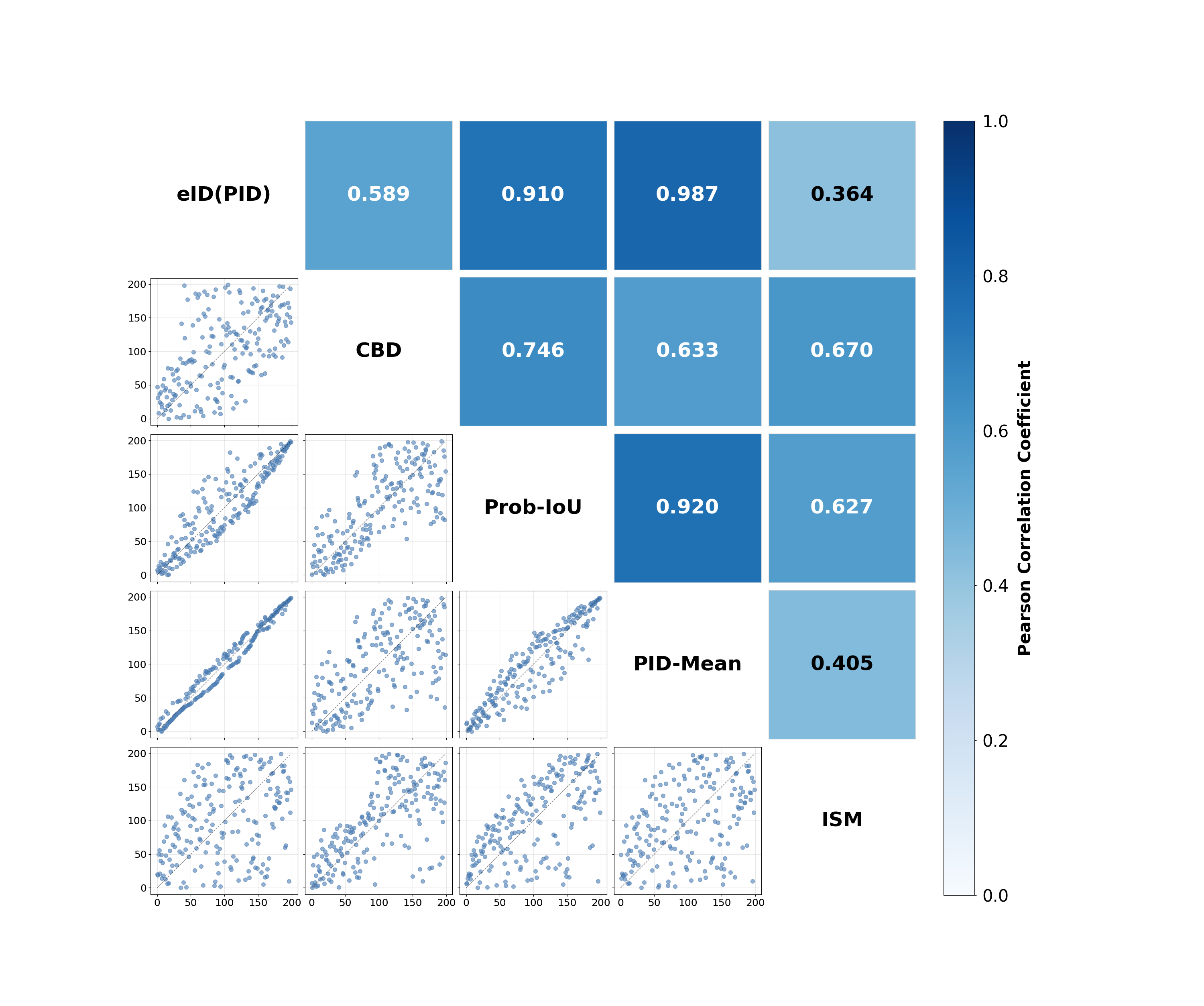}
    %\vspace{-1em}
    \caption{
    The scatterplot matrix shows ranking consistencies between five depth methods: eID (binary specialization of PID), CBD, probabilistic IoU depth (Prob-IoU), PID-mean, and isosurface similarity map depth (ISM), on synthetic contour ensembles. 
   Scatterplots in the lower triangle visualize pairwise depth rankings, where each point represents ranks of an ensemble member under two different methods; the corresponding Pearson correlation coefficients are shown in the upper triangle.}
     \label{fig:consistencyTest}
\end{figure}

The results are shown in \cref{fig:consistencyTest} as a scatterplot matrix. The lower triangle shows scatterplots of pairwise rankings of ensemble members under two different methods, while the upper triangle shows the corresponding Pearson correlation coefficients. 
The quantitative results indicate a strong consistency between PID-mean and eID (PID), with a Pearson correlation coefficient of 0.987, the highest among all pairs. The scatterplot between these two methods reveals that most points are concentrated along the diagonal line, confirming minimal ranking disagreements.
The agreement is particularly high for members dissimilar to the median (upper right region of the scatterplot), indicating robust consistency for outlier detection.
The high correlation value indicates that PID-mean preserves the essential center-outward ordering inherent in depth-based contour analysis. 
This validates PID-mean as a reliable approximation suitable for large-scale contour ensemble analysis.

Moreover, it is interesting to learn from~\cref{fig:consistencyTest} that the Prob-IoU depth shows relatively high consistencies with PID-mean (Pearson correlation coefficient of 0.920) and eID (0.910), whereas CBD and ISM exhibit low consistencies versus all depth methods. 

\begin{figure}[tb]% >>>
  \centering
  \begin{minipage}[t]{.495\linewidth}
    \subcaptionbox{Spaghetti plots of 50 members of the ensemble}
      {\includegraphics[width=\linewidth]{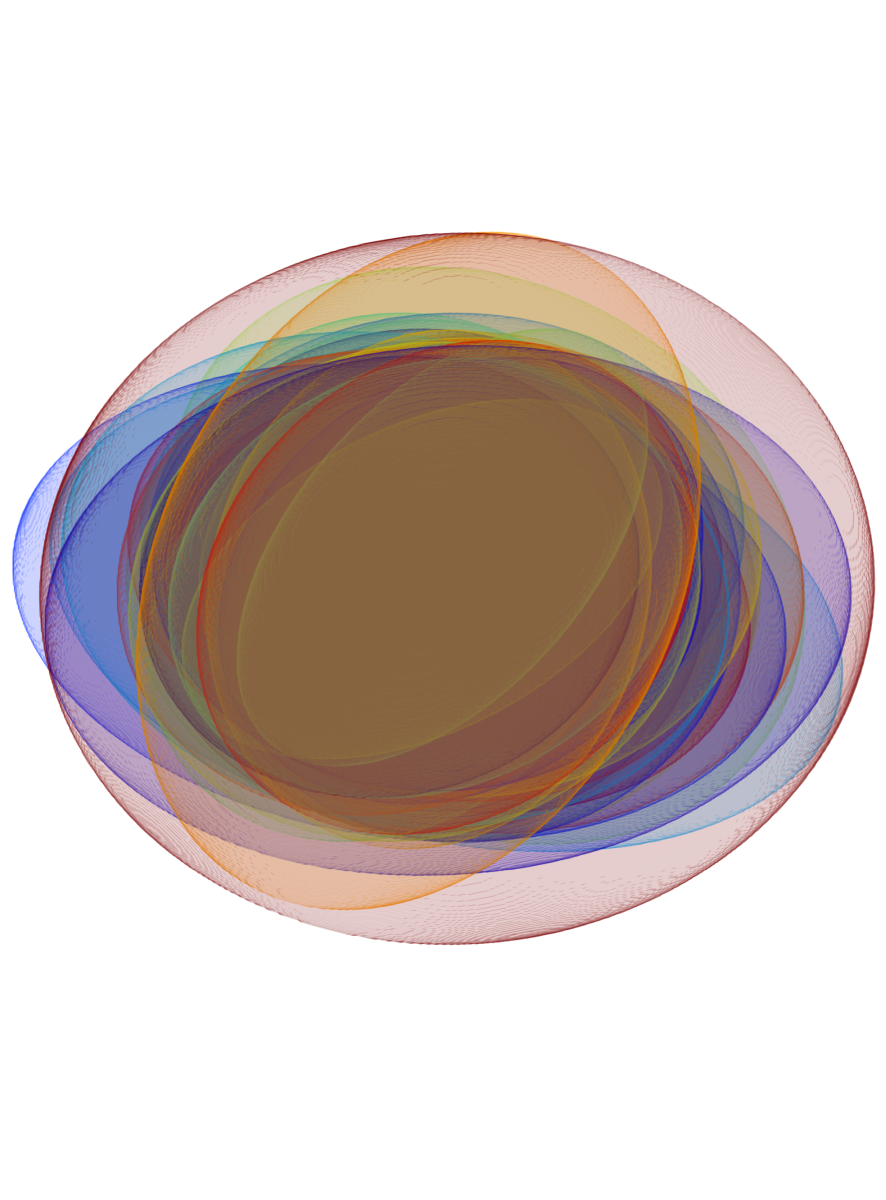}}%
  \end{minipage}%
  \hfill
  \begin{minipage}[b]{.5\linewidth}
    \subcaptionbox{Ensemble size scaling result}
      {\includegraphics[width=\linewidth]{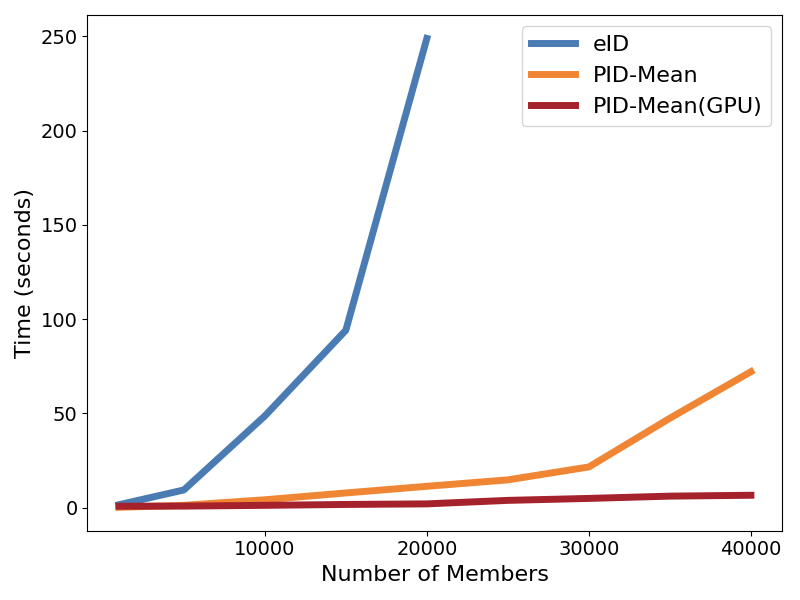}}
    \subcaptionbox{Data resolution scaling result}
      {\includegraphics[width=\linewidth]{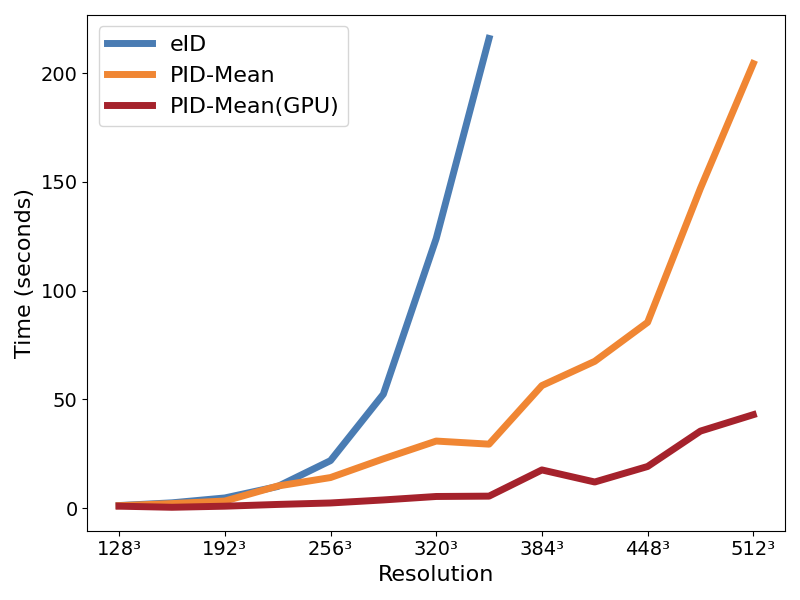}}%
  \end{minipage}%
   \caption{The scalability study comparing PID-mean, PID-mean (GPU), and eID uses 50 overlaid synthetic ellipsoid ensemble members visualized as (a) spaghetti plots. Scaling performance results are plotted for (b)~ensemble size and (c) data resolution. }
   \label{fig:3dsynth}
\end{figure}% <<<

\begin{figure*}[tb]
        \parbox[b]{.12\textwidth}{
     \subfloat[A slice of the T1 image]{\includegraphics[width=\linewidth]{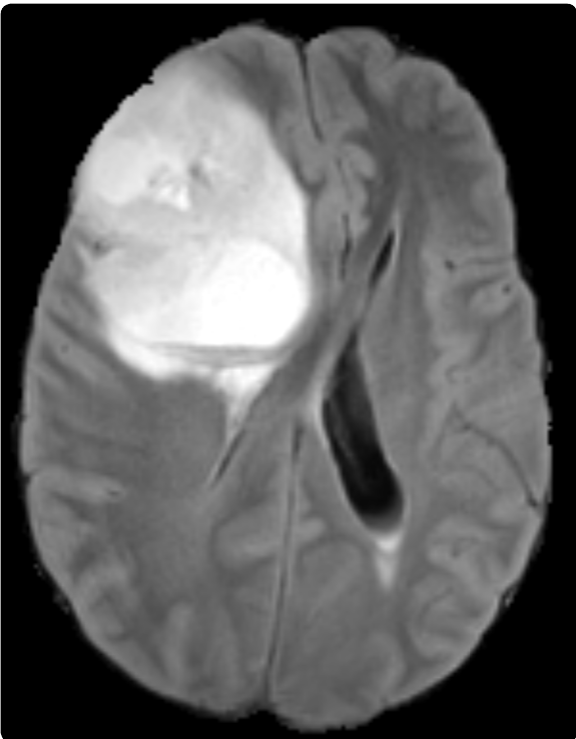}}\\
     \subfloat[A soft mask slice]{\includegraphics[width=\linewidth]{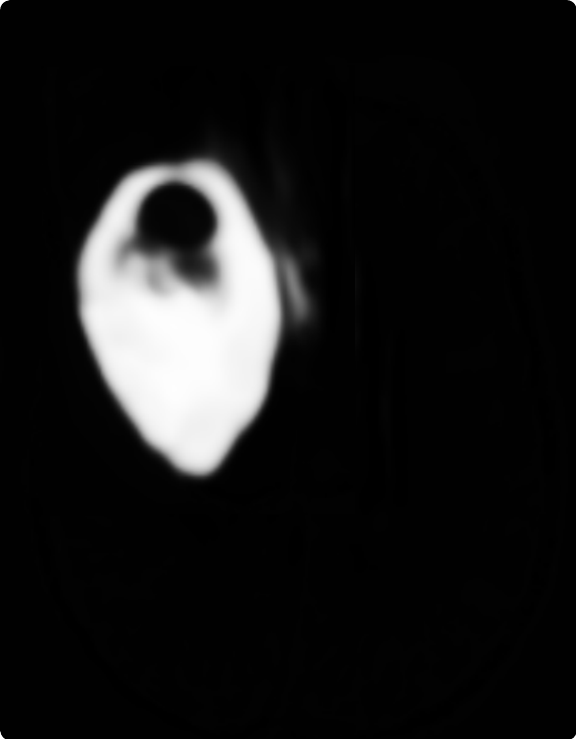}}
    }\hfill
    \parbox[b]{4.71cm}{
    \subfloat[Spaghetti plots of thresholded soft masks]{\includegraphics[width=\linewidth]{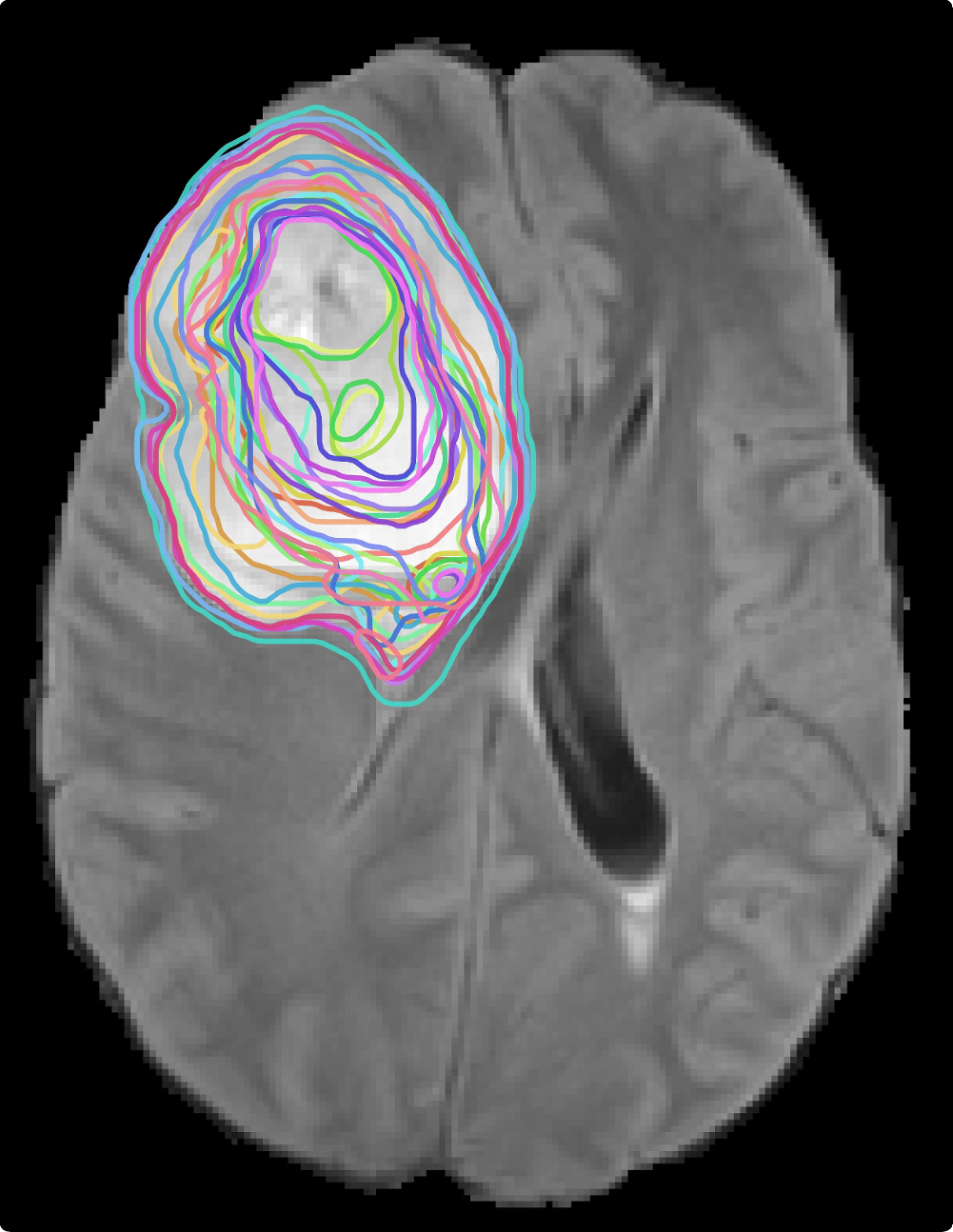}}
    }\hfill
    \parbox[b]{.55\textwidth}{\subfloat[3D contour boxplots]{\includegraphics[width=\linewidth]{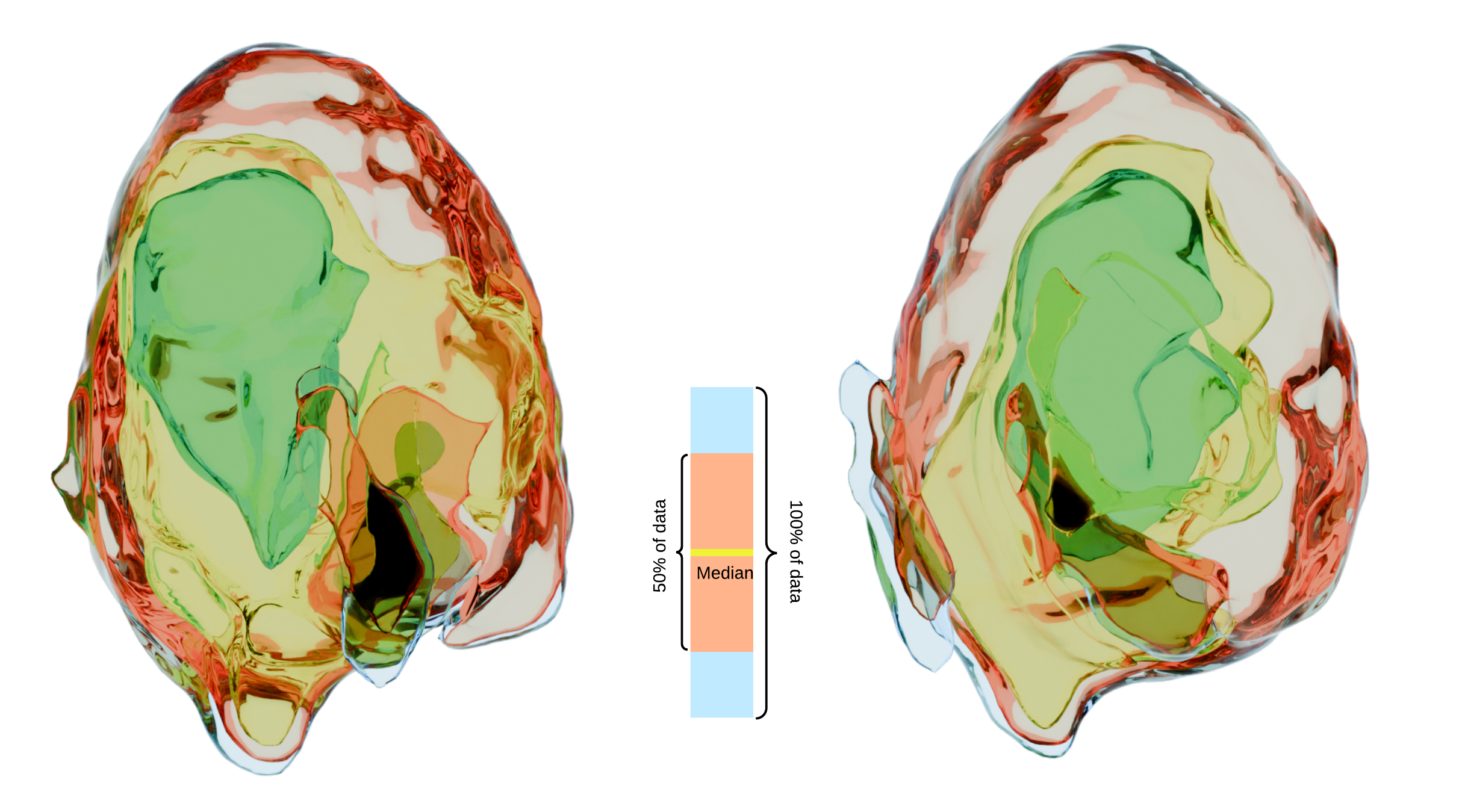}\vspace{0.8em}}}
    \vspace{-0.5em}
    \caption{Visualization of an ensemble of brain tumor soft masks from 31 segmentation networks. A slice of the T1 brain scan is shown in (a), and a soft mask slice of a member is shown in (b). Spaghetti plots of soft masks (with a threshold of probability=0.5 for good visibility) can be seen in (c). Two views of contour boxplots of the soft masks processed with PID (d) provide a summary visualization.   }
    \label{fig:braintumor}
\end{figure*}

\subsection{Scalability Study}
\label{sec:scalability}

Complementing the theoretical complexity analysis from \cref{sec:pidmean}, we also conducted an empirical study of computational costs with respect to varying problem size.
Here, we compare the eID, PID-mean, and PID-mean (GPU) implementations on 3D synthetic ellipsoid ensembles.

\paragraph{Experimental Setup}
We tested two scaling scenarios using synthetic 3D ellipsoid ensembles with controlled contamination (80\% normal samples, 20\% outliers); see \cref{fig:3dsynth}(a).
The first experiment scales the ensemble size from 100 to 40,000 masks with a fixed $50^3$ voxel data resolution, and the second experiment scales the data resolution from $128^3$ to $512^3$ voxels with a $32$ increment step size and with a fixed number of 50 ensemble members.

\paragraph{Performance Results}
All three methods exhibit super-linear scaling behavior in practice, with PID-mean achieving substantial speedup over eID, and PID-mean (GPU) having the highest performance improvements. 
\Cref{fig:3dsynth}(b) shows the ensemble size scaling results: for ensembles ranging from 10,000 to 20,000 members, eID takes 48.62--248.85 seconds, PID-mean completes in 4.17--11.32 seconds (11.7--22.0$\times$ speedup), while PID-mean (GPU) takes 1.23--2.00 seconds (39.6--124.4$\times$ speedup over eID, 3.4--5.7$\times$ speedup over PID-mean). 
The data resolution scaling performance is shown in \cref{fig:3dsynth}(c): for resolutions from 256$^3$ to 352$^3$ voxels, eID requires 21.84--216.03 seconds, PID-mean completes in 14.05--29.42 seconds (1.6--7.3$\times$ speedup), while PID-mean (GPU) achieves 2.36--5.50 seconds (9.3--39.3$\times$ speedup over eID, 5.3--6.0$\times$ speedup over PID-mean).
Notably, eID encounters memory limitations beyond 20,000 ensemble members and resolutions of 352$^3$ voxels, making it unable to process larger datasets, while both PID-mean variants continue to scale efficiently.

Overall, PID-mean (GPU) achieves a speedup of one to two orders of magnitude over eID. 
Even without GPU acceleration, PID-mean provides notable speedup over eID, making both variants suitable for analyzing 3D ensembles.

We confirm that PID-mean has high ranking consistency with eID, a special case of PID, while PID-mean offers significant speed and memory advantages over eID. 
Moreover, the parallel implementation further accelerates PID-mean.

\newcommand{\heightHipp}{3.4cm}

\begin{figure*}[tb]
    \centering

     \subfloat[Spaghetti plots of hippocampus on 2D slices (20 members)]{\includegraphics[height=\heightHipp]{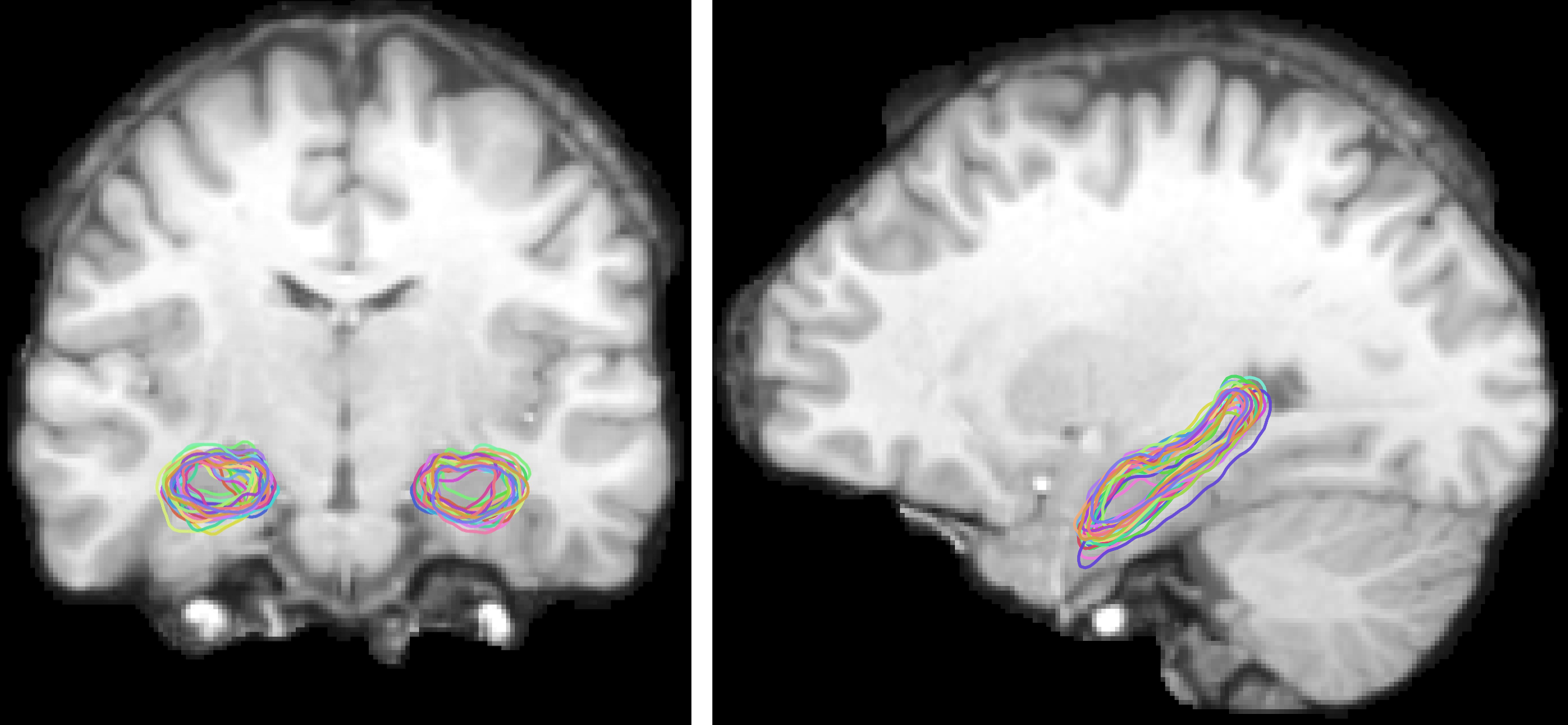}}\hfill
     \subfloat[3D spaghetti plots of hippocampus]{\includegraphics[height=\heightHipp]{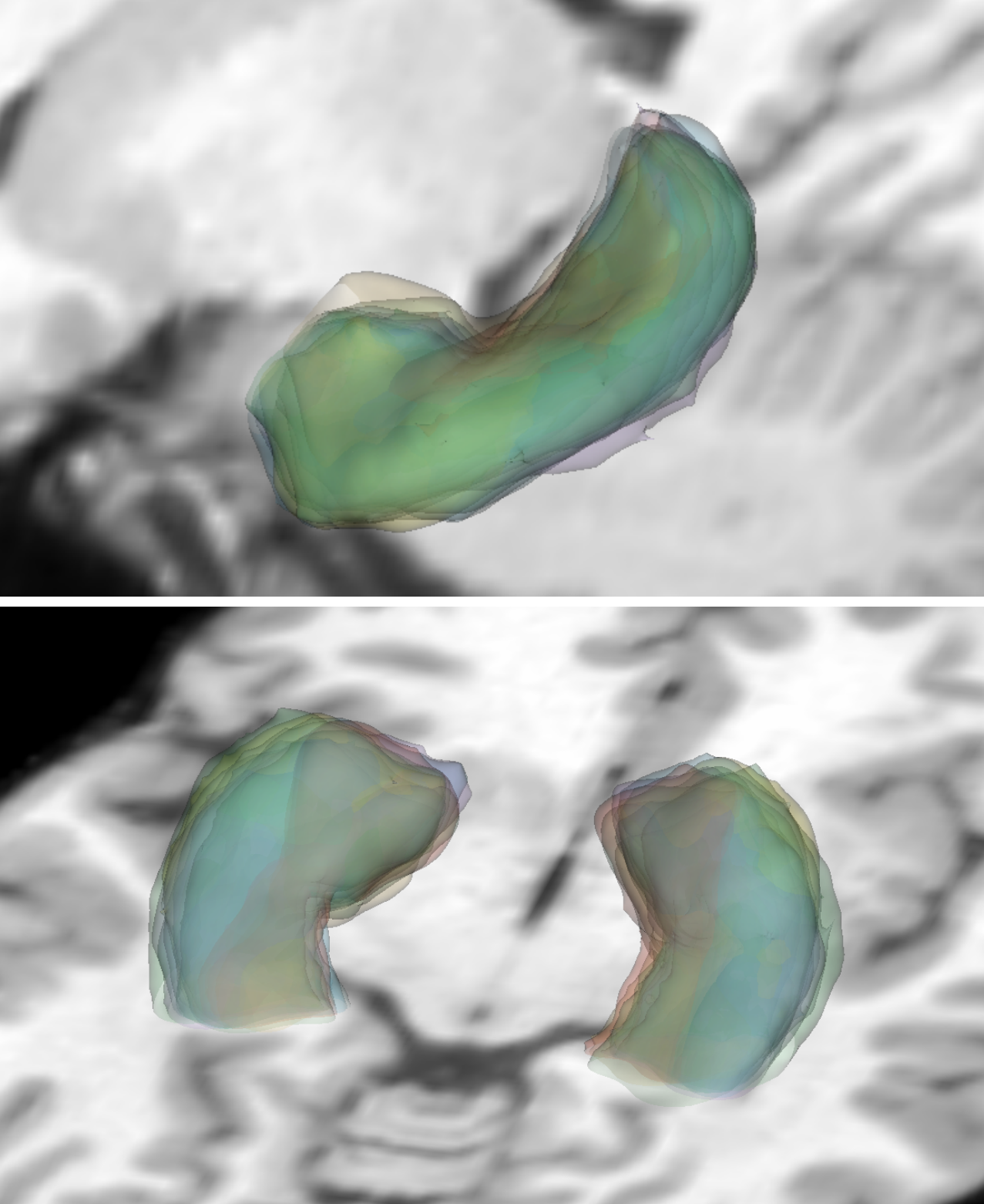}}
     \subfloat[3D contour plots of hippocampus]{\includegraphics[height=\heightHipp]{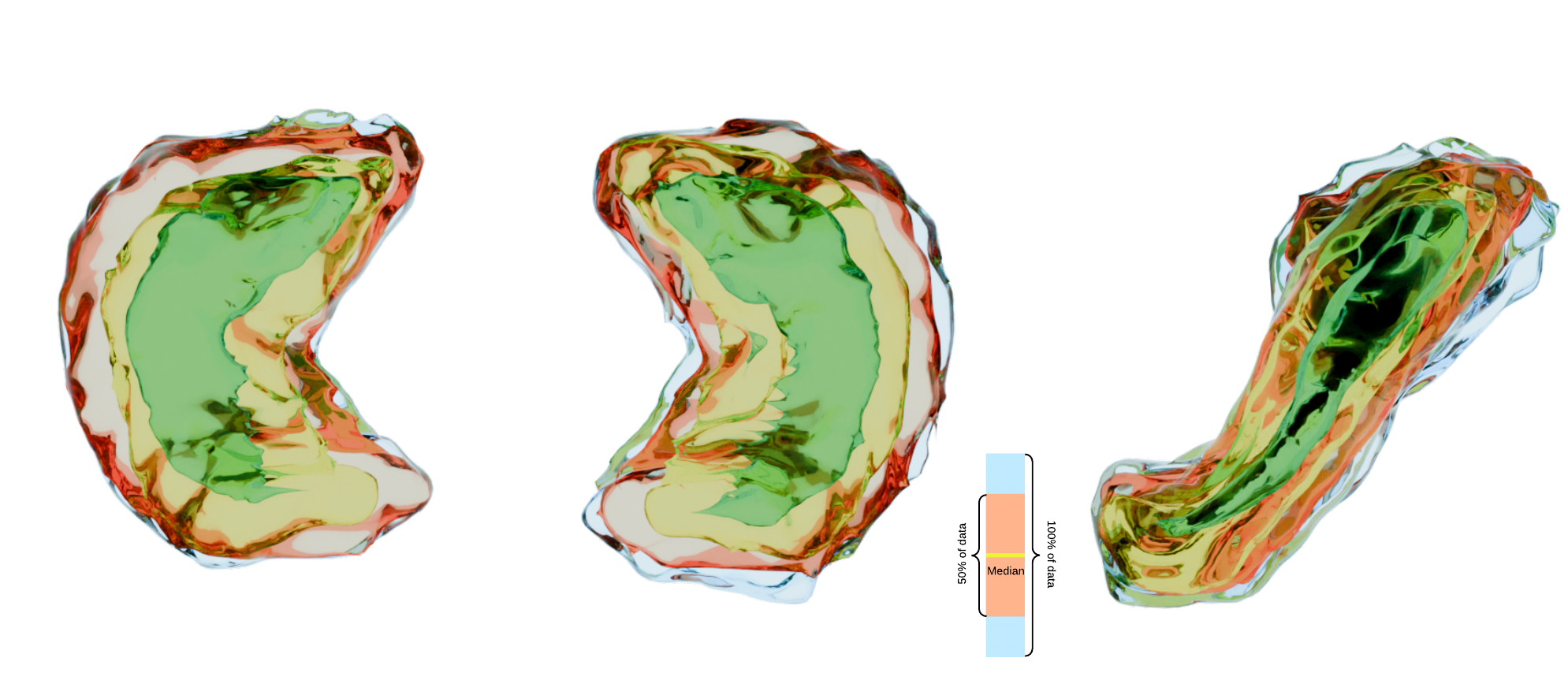}}
     \\
     \subfloat[Spaghetti plots of ventricles on 2D slices (20 members)]{\includegraphics[height=\heightHipp]{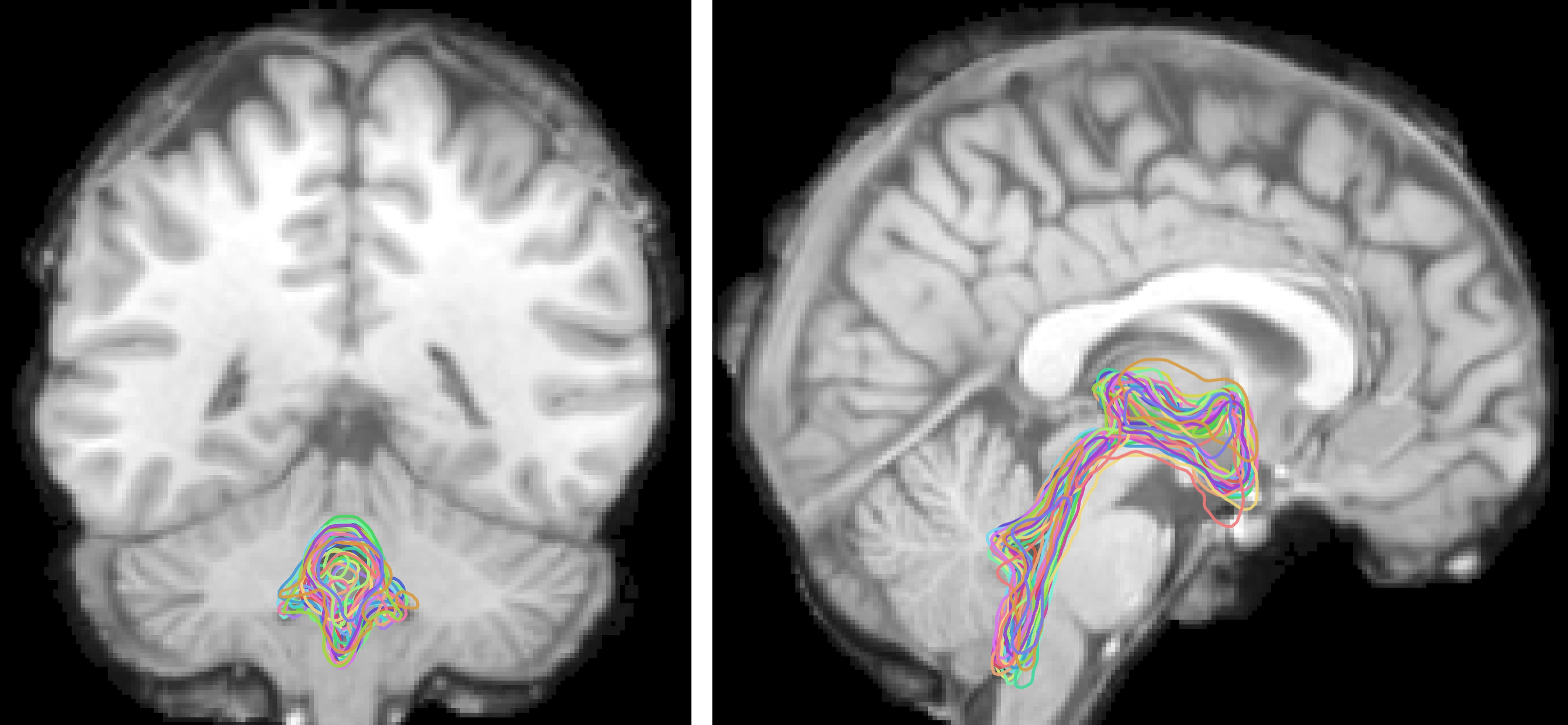}}
     \hfill
     \subfloat[3D spaghetti plots of ventricles]{\includegraphics[height=\heightHipp]{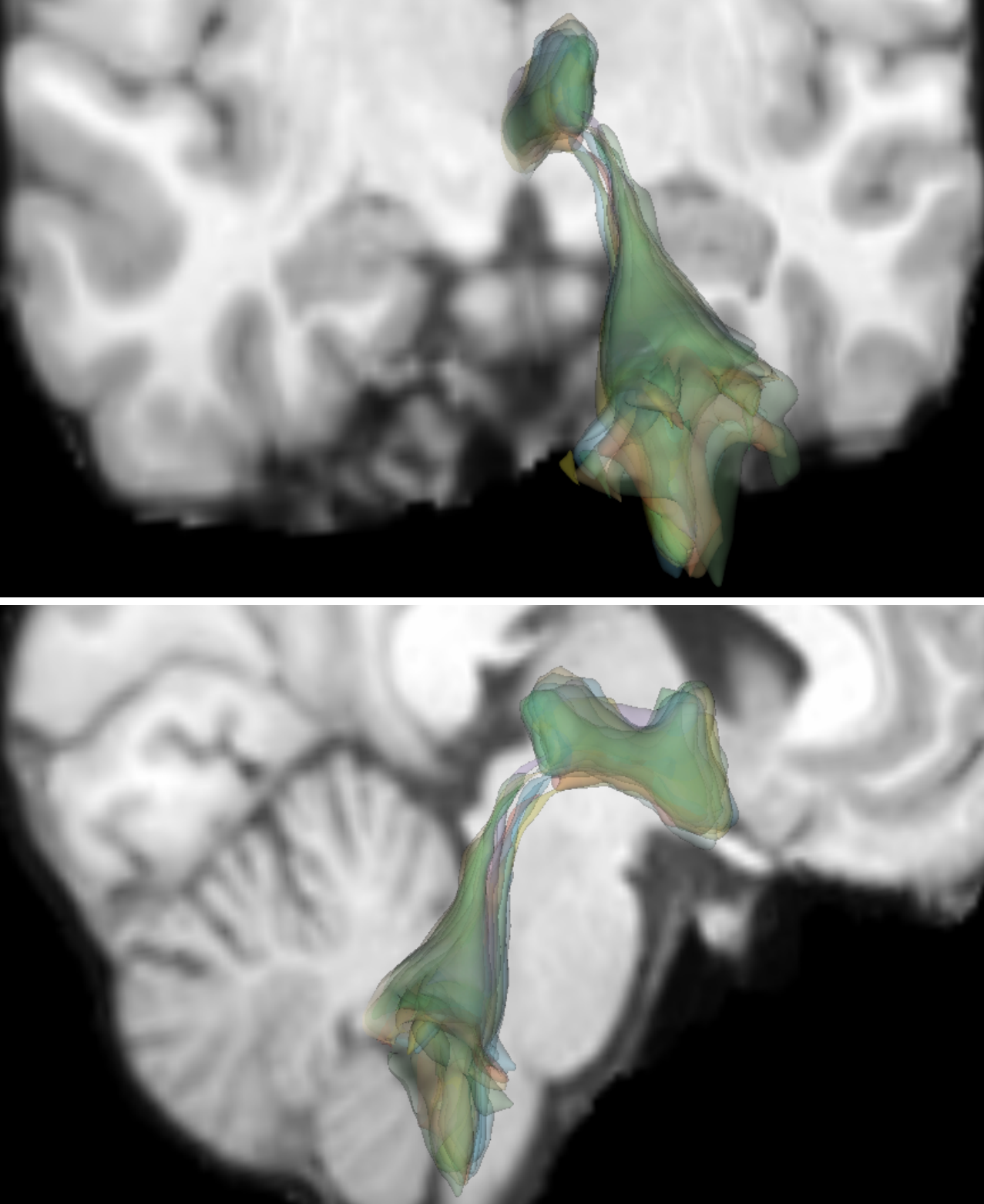}}
     \subfloat[3D contour plots of ventricles]{\includegraphics[height=\heightHipp]{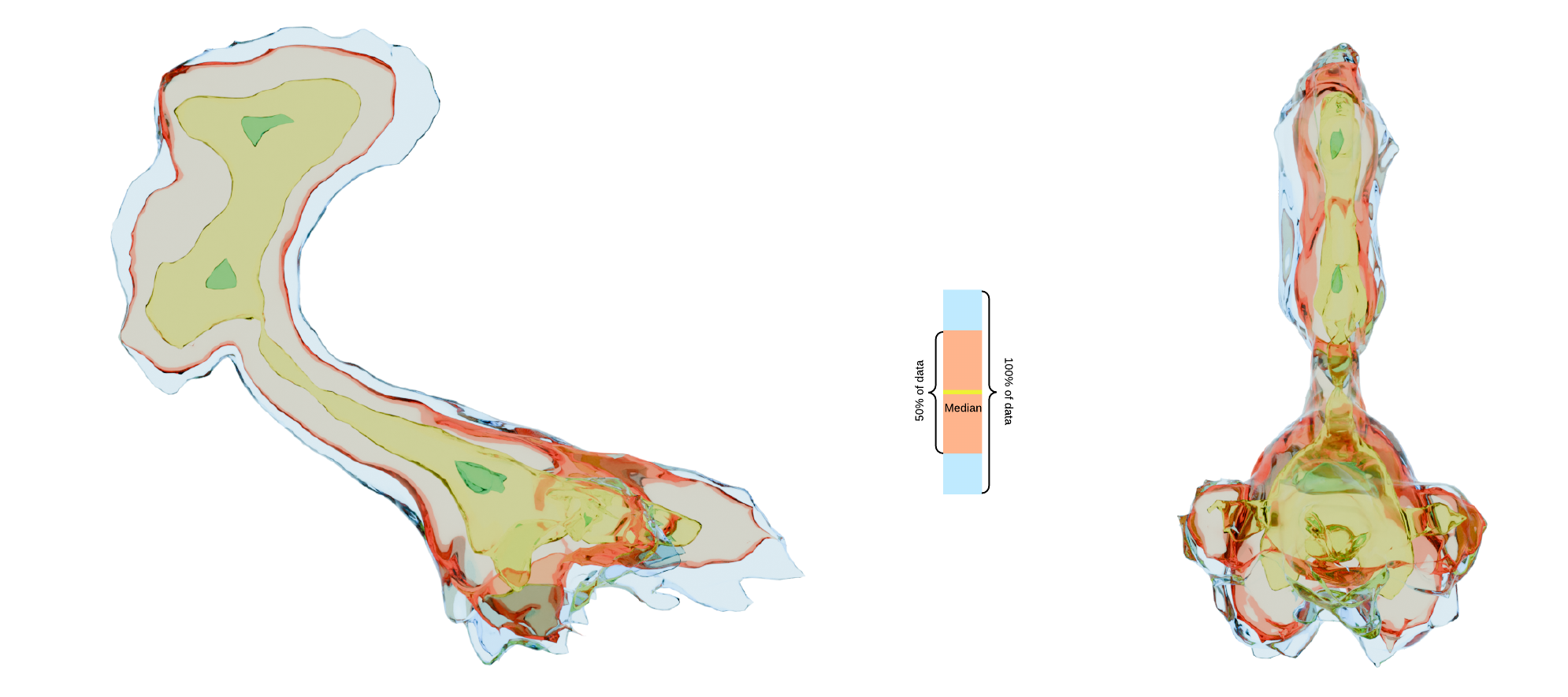}}

    \caption{Visualizations of ensembles of 400 segmented hippocampus and third and fourth ventricles of the IXI dataset. Spaghetti plots of hippocampi and ventricles of a few members on (a, d) 2D slices and (b, e) in 3D are already cluttered. In contrast, 3D contour boxplots processed by PID in (c) and (f) give much clearer visualizations.  }
    \label{fig:ixidataset}
\end{figure*}

\section{Examples}
The effectiveness of our method is demonstrated through several real-world 3D ensembles.  
These examples cover various data types to show the usefulness of our method, including ensembles of traditional binary masks, soft masks for segmentation with neural networks trained with different parameters, and fuzzy surfaces from scalar field ensembles.

Data depths of all examples are computed by GPU-accelerated PID-mean and used to generate 3D contour boxplots. 
The 3D contour boxplots are computed by taking the unions and intersections of surfaces (binarization is needed for probabilistic maps after the PID-mean processing for visualization) of populations in certain percentiles to form corresponding envelopes. 
Refraction effects are included to visualize the spatial complexity of the contour boxplots, as refraction can provide important shape cues for human perception, which is particularly useful for our case of nested transparent structures~\cite{Magnus2018}.
High-quality renderings with refractions are achieved with global illumination in Blender.
Figures of larger sizes and videos of the refraction rendering results can be found in the supplemental material~\cite{supplemental:ProbInclDepth}.

\subsection{Brain MRI Segmentation Soft Masks}
\label{sec:brainTumor}
Our method is applied to ensembles of soft masks from a collection of segmentation models, as shown in \cref{fig:braintumor}.
We use a 3D brain tumor MRI dataset (T1/T1ce/T2/FLAIR)---MSD Tumour~\cite{antonelli2022medical}---with voxel-wise labels for four classes (background, necrotic/non-enhancing core, edema, enhancing tumor), as shown in \cref{fig:braintumor}(a). The data has a spatial resolution of $224 \times 224 \times 144$ voxels.
For the ensemble input to PID-mean, we train 31 SegResNet models under different initialization and training-stabilization configurations while keeping the architecture and augmentations fixed, then collect the soft mask outputs (one exemplary soft mask is shown in \cref{fig:braintumor}(b)) of these 31 models on the same test data (\cref{fig:braintumor}(c)). 
Details of the data preparation are covered in the supplemental material~\cite{supplemental:ProbInclDepth}.

Contour boxplots of 3D soft mask ensembles (thresholded at a probability of 0.5 to create surfaces) are sorted by PID-mean and visualized in \cref{fig:braintumor}(d). 
Refractions highlight fine details of the complex shapes of the brain tumor.
It can be seen that the envelopes of 50\% members (orange) and 100\% (blue) members are largely identical but differ in several locations on the outer boundaries, while the median member (yellow) is inside the band of 50\% members.
Due to color blending, the inner surface of the 100\% members band in blue appears light emerald green, and this holds for other examples in the section too.

The contour boxplot can help with segmentation quality control for soft masks.
Low-depth contours identified by PID-mean can effectively flag high-risk or outlier segments for review,
whereas high-depth contours indicate the typical segment from the collection of segmentation networks.
By further examining the associated soft masks, one can improve the settings of the segmentation model for desired results.

\subsection{IXI Dataset}
\label{sec:ixidata}

Our method can be used to effectively process 3D binary masks using the binary specialization of GPU PID-mean. 
Examples are shown for the publicly available Information eXtraction from Images (IXI) dataset~\cite{antonelli2022medical}\footnote{\url{https://brain-development.org/ixi-dataset/}}. 
Specifically, we use 400 T1-weighted MRI volumes of size $160\times224\times192$. 

For structure-specific analysis, we consider the segmentation labels of the hippocampus (\cref{fig:ixidataset}(a, b, c)) and the third and fourth ventricles (\cref{fig:ixidataset}(d, e, f)). 
Binary contours of a few members are visualized as spaghetti plots on 2D slices (\cref{fig:ixidataset}(a, d)) and as 3D isosurfaces (\cref{fig:ixidataset}(b, e)).
These spaghetti plots reveal the fine structures of the contours but are already cluttered, especially for the 3D cases, which suffer from severe occlusions.

Clear visualizations are achieved with 3D contour boxplots generated from the PID-mean outputs. 
Contour boxplots of the hippocampus (\cref{fig:ixidataset}(c)) reveal the high agreement between the envelope of 50\% (orange) members and 100\% members (blue), and the median member (yellow) shows the representative shape and size of the ensemble.
While for the ventricles, contour boxplots (\cref{fig:ixidataset}(f)) show that there is visible space between the envelope of the 50\% members and that of the 100\% members, the median member reveals the typical ventricles of the ensemble.
Regions inside the interior surfaces are small, indicating that variations between members are rather large. 

PID-mean helps gain insights into the trends of complex anatomical structures within the binary contour ensemble.
The example demonstrates that PID-mean has the potential for analyzing specific medical problems in cohort studies using ensembles of medical images.

\subsection{ScalarFlow}
\label{sec:scalarFlow}
\begin{figure}[tb]
    \centering
    \includegraphics[width=0.9\linewidth]{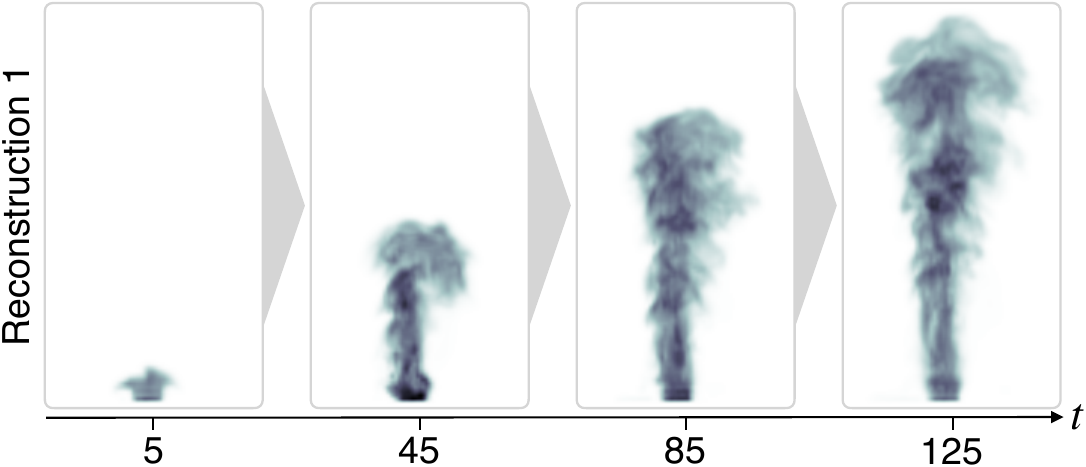}
    \caption{Visualization of the first reconstruction in the ScalarFlow dataset at time steps 5, 45, 85, and 125. The whole set of 3D reconstructions is used for the PID-mean computation and the boxplot view in \cref{fig:teaser}.}
    \label{fig:ScalarFlow_1}
\end{figure}

ScalarFlow~\cite{ScalarFlow2019}, a large-scale volumetric dataset of real-world smoke plumes captured with a multi-view setup and reconstructed via a physics-based, simulation-constrained tomography framework, is used as a third example.
The data comprise time-resolved volumetric fields with buoyancy-driven flows that transition to turbulence, together with camera calibrations and accompanying image sequences. 

We process all time steps of the density volumes (selected time steps of one member are shown in \cref{fig:ScalarFlow_1}) to show the effectiveness of GPU-based PID-mean on 3D scalar fields.
We model the normalized density fields as probabilistic maps that describe the probability of the appearance of smoke plumes, i.e., higher normalized density indicates a higher chance of appearance.

Contour boxplots of four representative time steps (thresholded by a probability of 0.02 to create surfaces) are shown in~\cref{fig:teaser} (Box Plot View).
Early time steps ($t=5, 45$) exhibit comparatively simple structures and high agreement between 50\% and 100\% members as seen in the zoom-ins.
Especially, the very early 5th time step shows low-relief surfaces.
Later time steps ($t=85, 125$) reveal highly complex and more turbulent smoke plume structures, as shown by the delicate surfaces and dark regions resulting from complicated refractions. The differences between 50\% and 100\% members are much higher than those in earlier steps, indicating the high uncertainty of smoke after time progression.
The boxplot views reflect the complexity of the underlying smoke plume reconstructions. 
This probabilistic overview of ensemble member densities enables an interpretable and data-driven analysis of the whole dataset at a glance.
Furthermore, the reconstructions do not deviate completely but match the expected formation of smoke.

PID is dimension-agnostic and can be applied to 4D spatiotemporal ensembles by treating time as an additional dimension.
This yields a global depth ranking that accounts for spatiotemporal coherence across the entire dynamic process.
We demonstrate this capability on the full 4D ScalarFlow ensemble in the supplemental material~\cite{supplemental:ProbInclDepth}.

These examples of complex 3D ensembles are visualized as 3D contour boxplots thanks to the efficient processing by GPU-accelerated PID-mean.
These qualitative results suggest that the uncertainty-aware nature of PID-mean/PID facilitates the analysis of data with fuzzy contours either from explicit soft masks (\cref{sec:brainTumor}) or by modeling a probability distribution of scalar fields (\cref{sec:scalarFlow}).
The example of binary contours (\cref{sec:ixidata}) shows that our method can be used as an alternative to existing data depth methods for its speed benefits.

\section{Expert Interview}
Expert feedback on our method was obtained through interviews.
Since our method is introduced in the context of medical, fuzzy, and scientific data, we interviewed three experts (each 5+ years of experience in at least one of the fields). In the following, we refer to the experts using the abbreviation for their key expertise (M, F, and S).
The experts volunteered to be part of the interview and had no access to the content of this paper.
To introduce our method, we presented each expert individually with descriptions of data depth, our PID computation, and the results. Thereby, we used the illustrations from the paper.
We prepared three questions on 1) how well the visualization summarizes the ensemble, 2) the clarity of meanings of surfaces, and 3) specific scenarios where PID could be beneficial, and left space for an open discussion afterward.
Each interview lasted between 30 and 45 minutes. 

All three experts agree that the visualization provides adequate ensemble summarization (F: ``Yeah, if you want to look at ensembles'') with challenges in the visual complexity. M noted that the effectiveness could vary by application domain, e.g., it could be misleading for variable pathologies like tumors.
The conceptual meanings of the different surfaces were generally well-received across all experts and described as ``clear'' and ``intuitive.'' Only M noted: for ``doctor's understanding, some more explanations may be needed.''

Asking for specific scenarios where the technique could be beneficial, F noted that ``if you want to look at ensembles'' you could ``get an overview,'' but it would be interesting to enable an identification of outliers. S noted that the method probably ``works best with clean simulation data rather than noisy experimental'' data and M mentioned the use case of tracking disease progression for patients.
In the open discussion, the experts remarked that interactively changing the shown contour boxplots would be generally very beneficial for using the method ``because you always have occlusion in 3D'' (M).

\section{Conclusion and Future Work}
We have presented PID for visualizing fuzzy contours extracted from ensembles of scalar fields.
PID integrates the probabilistic distribution of contours into depth computation, allowing the analysis of scalar field ensembles and soft masks generated by segmentation methods. 
The combination of PID-mean and the parallel implementation makes depth analysis of high-resolution 3D ensembles feasible.
Our method is evaluated by ranking consistency tests and scalability analysis, validating the correctness and efficiency of our method. 
The usefulness of our method is further demonstrated using real-world 3D ensembles of various types from several application areas.
Our method is a generalized model for contour ensembles and can process binary contours as a specialization.
While data depth for fuzzy contours helps visualize the sensitivity of the scalar fields, data depth of binary contours faithfully describes specific binary decisions. 
Therefore, we argue that both cases are valuable for gaining insights into an ensemble as they complement each other---the fuzzy contours depict the global aspect while binary contours describe the local aspect of an ensemble.

One limitation of our method is the rendering of fuzzy surfaces.
Currently, the surface rendering limits us to visualize binarized representatives for fuzzy surfaces. This binarization for visualization introduces data-dependent sensitivity when choosing the representative for, mainly, the median surface.
However, it is important to note that, unlike existing depth methods for binary contours, the full probabilistic information is preserved in our method and can be used for visualization.

For future work, we would like to study the visual representation of ranked fuzzy contours. 
We aim to devise perceptually enhanced glyph renderings that encode the fuzziness on 3D surfaces to visualize the full information, or allow users to interactively browse the range of fuzzy contours.
Another direction is to support interactive analysis on user-selected regions of an ensemble, as analysts are often interested in local rather than global data structures in space and the behaviors of members thereof.  
Finally, we would like to extend PID for spatiotemporal probabilistic surfaces and non-spatial data such as dynamic~graphs.

%% ---------------------------------------------------------------------------

%% if specified like this the section will be omitted in review mode
\acknowledgments{%
  This work was supported in part by a National Science Foundation of China grant (62372012), and the Deutsche Forschungsgemeinschaft (DFG, German Research Foundation) -- Project-ID 251654672 -- TRR~161.%
}

\bibliographystyle{abbrv-doi-hyperref}

\bibliography{datadepth}

\end{document}